\documentclass[english,aps,prd,nofootinbib,showkeys,preprint,floatfix]{revtex4}

\usepackage{amsmath}
\usepackage{amssymb}
\usepackage{graphicx}
\usepackage{rotating}
\usepackage{color} 
\usepackage{multirow} 
\usepackage[T1]{fontenc} 
\usepackage[latin9]{inputenc}
\usepackage{lmodern}
\makeatletter
\providecommand{\tabularnewline}{\\}
\usepackage{booktabs}
\usepackage{mathrsfs}   
\usepackage{slashed}     
\usepackage{url}
\usepackage{multirow}
\usepackage{units}
\usepackage{ wasysym }

\makeatother

\usepackage{babel}

\newcommand{\AddrAHEP}{
  {\it AHEP Group, Instituto de F\'{\i}sica Corpuscular --
    C.S.I.C./Universitat de Val{\`e}ncia \\
    Edificio de Institutos de Paterna, Apartado 22085,
  E--46071 Val{\`e}ncia, Spain}}

\newcommand{\AddrLisb}{%
 Departamento de F\'\i sica and CFTP, Instituto Superior T\'ecnico\\
          Av. Rovisco Pais 1, 1049-001 Lisboa, Portugal }

\def\gsim{\raise0.3ex\hbox{$\;>$\kern-0.75em\raise-1.1ex\hbox{$\sim\;$}}}
\def\lsim{\raise0.3ex\hbox{$\;<$\kern-0.75em\raise-1.1ex\hbox{$\sim\;$}}}

\begin{document}

\preprint{CFTP/13-002}  
\preprint{IFIC/13-02}  
\hfill \today

\title{Supersymmetric $SO(10)$ GUTs with sliding scales}

\author{Carolina Arbel\'aez}\email{Carolina.Arbelaez@ist.utl.pt}
\author{Renato M. Fonseca }\email{renato.fonseca@ist.utl.pt}
\author{Jorge C. Rom\~ao}\email{jorge.romao@ist.utl.pt}
\affiliation{\AddrLisb} 
\author{Martin Hirsch} \email{mahirsch@ific.uv.es}
\affiliation{\AddrAHEP}



\begin{abstract}
We construct lists of supersymmetric models with extended gauge groups
at intermediate steps, all of which are based on SO(10)
unification. We consider three different kinds of setups: (i) The
model has exactly one additional intermediate scale with a left-right
(LR) symmetric group; (ii) SO(10) is broken to the LR group via an
intermediate Pati-Salam (PS) scale; and (iii) the LR group is broken
into $SU(3)_{c} \times SU(2)_{L} \times U(1)_{R} \times U(1)_{B-L}$,
before breaking to the SM group.  We use sets of conditions, which we
call the ``sliding mechanism'', which yield unification with the
extended gauge group(s) allowed at arbitrary intermediate energy
scales. All models thus can have new gauge bosons within the reach of
the LHC, in principle. We apply additional conditions, such as
perturbative unification, renormalizability and anomaly cancellation
and find that, despite these requirements, for the ansatz (i) with
only one additional scale still around 50 different variants exist
that can have an LR symmetry below 10 TeV. For the more complicated
schemes (ii) and (iii) literally thousands of possible variants exist,
and for scheme (ii) we have also found variants with very low PS
scales.  We also discuss possible experimental tests of the models
from measurements of SUSY masses.  Assuming mSugra boundary conditions
we calculate certain combinations of soft terms, called
``invariants'', for the different classes of models. Values for all
the invariants can be classified into a small number of sets, which
contain information about the class of models and, in principle, the
scale of beyond-MSSM physics, even in case the extended gauge group is
broken at an energy beyond the reach of the LHC.

\end{abstract}

\maketitle

\section{Introduction}

In the MSSM (``Minimal Supersymmetric extension of the Standard
Model'') gauge couplings unify at an energy scale of about $m_G \simeq
2 \times 10^{16}$ GeV. Adding particles arbitrarily to the MSSM 
easily destroys this attractive feature. Thus, relatively few 
SUSY models have been discussed in the literature which have a 
larger than MSSM particle content at experimentally accessible 
energies. Neutrino oscillation experiments
\cite{Fukuda:1998mi,Ahmad:2002jz,Eguchi:2002dm}, however, have shown that at
least one neutrino must have a mass $m_{\rm Atm} \ge 0.05$ eV. 
\footnote{For the latest fits of oscillation data, see for example
\cite{Tortola:2012te}.} A (Majorana) neutrino mass of this order 
indicates the existence of a new energy scale below $m_G$. For 
models with renormalizable interactions and perturbative couplings, 
as for example in the classical seesaw models 
\cite{Minkowski:1977sc,Yanagida:1979ss,GellMann:1980vs,Mohapatra:1979ia}, 
this new scale should lie below approximately $\Lambda_{\rm LNV} \lsim 
10^{15}$ GeV.

From the theoretical point of view GUT models based on the group
$SO(10)$ \cite{Fritzsch:1974nn} offer a number of advantages compared
to the simpler models based on $SU(5)$. For example, several of the
chains through which $SO(10)$ can be broken to the SM gauge group
contain the left-right symmetric group $SU(3)_{c} \times SU(2)_{L}
\times SU(2)_{R} \times U(1)_{B-L}$ as an intermediate step 
\cite{Mohapatra:1986uf}, thus potentially explaining the observed
left-handedness of the weak interactions. However, probably the most
interesting aspect of $SO(10)$ is that it automatically contains the
necessary ingredients to generate a seesaw mechanism
\cite{Mohapatra:1979ia}: (i) the right-handed neutrino is
included in the ${\bf 16}$ which forms a fermion family;
and (ii) $(B-L)$ is one of the generators of $SO(10)$.

Left-right (LR) symmetric models usually break the LR symmetry at a
rather large energy scale, $m_R$. For example, if LR is broken in the
SUSY LR model by the vev of $(B-L)=2$ triplets
\cite{Cvetic:1983su,Kuchimanchi:1993jg} or by a combination of
$(B-L)=2$ and $(B-L)=0$ triplets \cite{Aulakh:1997ba,Aulakh:1997fq},
$m_R \simeq 10^{15}$ GeV is the typical scale consistent with gauge
coupling unification (GCU). The authors of \cite{Majee:2007uv} find a
lower limit of $m_R \gsim 10^9$ GeV from GCU for models where the LR
symmetry is broken by triplets, even if one allows additional
non-renormalizable operators or sizeable GUT-scale thresholds to be
present. On the other hand, in models with an extended gauge group it
is possible to formulate sets of conditions on the $\beta$-coefficients for
the gauge couplings, which allow to enforce GCU independent of the
energy scale at which the extended gauge group is broken. This was
called the ``sliding mechanism'' in \cite{DeRomeri:2011ie}.
\footnote{A different (but related) approach to enforcing GCU is taken
by the authors of \cite{Calibbi:2009cp} with what they call ``magic
fields''.} However, \cite{DeRomeri:2011ie} was not the
first to present examples of ``sliding scale'' models in the literature. 
In \cite{Malinsky:2005bi} it was shown that, if 
the left-right
group is broken to $SU(2)_L\times U(1)_R \times U(1)_{B-L}$ by the
vacuum expectation value of a scalar field $\Phi_{1,1,3,0}$ then 
\footnote{The indices are the transformation properties under the LR
group, see next section and appendix for notation.}  the resulting
$U(1)_R\times U(1)_{B-L}$ can be broken to $U(1)_Y$ of the SM in
agreement with experimental data at any energy scale. In
\cite{Majee:2007uv} the authors demonstrated that in fact a complete
LR group can be lowered to the TeV-scale, if certain carefully chosen
fields are added and the LR-symmetry is broken by right doublets. A
particularly simple model of this kind was discussed in
\cite{Dev:2009aw}.  Finally, the authors of \cite{DeRomeri:2011ie}
discussed also an alternative way of constructing a sliding LR scale
by relating it to an intermediate Pati-Salam stage. We note in passing
that these papers are not in contradiction with the earlier work
\cite{Cvetic:1983su,Kuchimanchi:1993jg,Aulakh:1997ba,Aulakh:1997fq},
which all have to have large $m_R$.  As discussed briefly in the next
section it is not possible to construct a sliding scale variant for an
LR model including pairs of $\Phi_{1,1,3,-2}$ and $\Phi_{1,3,1,-2}$.

Three different constructions, based on different $SO(10)$ breaking
chains, were considered in \cite{DeRomeri:2011ie}.  In chain-I
SO(10) is broken in exactly one intermediate (LR symmetric) step to
the standard model group: 
\begin{equation}\label{eq:chainI}
SO(10)\to SU(3)_{c} \times SU(2)_{L} \times SU(2)_{R}  \times U(1)_{B-L} 
\to \hskip2mm {\rm MSSM} .
\end{equation}
In chain-II SO(10) is broken first to the Pati-Salam group: 
\cite{Pati:1974yy} 
\begin{eqnarray}\label{eq:chainII}
SO(10) &\to & SU(4) \times SU(2)_{L} \times SU(2)_{R} \\ \nonumber
   &\to & SU(3)_{c} \times SU(2)_{L} \times SU(2)_{R} \times
  U(1)_{B-L} \to \hskip2mm {\rm MSSM} .
\end{eqnarray}
And finally, in chain-III:
\begin{eqnarray}\label{eq:chainIII}
SO(10) & \to & SU(3)_{c} \times SU(2)_{L} \times SU(2)_{R} \times U(1)_{B-L} 
\\ \nonumber
      & \to &  SU(3)_{c} \times SU(2)_{L} \times U(1)_{R} \times U(1)_{B-L}
        \to  \hskip2mm {\rm MSSM}. 
\end{eqnarray}
In all cases the last symmetry breaking scale before reaching the SM
group can be as low as ${\cal O}(1)$ TeV maintaining nevertheless
GCU. \footnote{In fact, the sliding mechanism would work also at even
lower energy scales. This possibility is, however, excluded
phenomenologically.} The papers discussed above
\cite{Malinsky:2005bi,Majee:2007uv,Dev:2009aw,DeRomeri:2011ie} give at
most one or two example models for each chain, i.e.  they present a
``proof of principle'' that models with the stipulated conditions
indeed can be constructed in agreement with experimental
constraints. It is then perhaps natural to ask: How unique are the
models discussed in these papers? The answer we find for this question
is, perhaps unsurprisingly, that a huge number of variants exist in
each class. Even in the simplest class (chain-I) we have found a total
of 53 variants (up to 5324 ``configurations'', see next section) which
can have perturbative GCU and a LR scale below 10 TeV, consistent with
experimental data.  For the two other classes, chain-II and chain-III,
we have found literally thousands of variants.

With such a huge number of variants of essentially ``equivalent''
constructions one immediate concern is, whether there is any way of
distinguishing among all of these constructions experimentally.  Tests
could be either direct or indirect. Direct tests are possible, because
of the sliding scale feature of the classes of models we discuss, see
section \ref{sec:models}. Different variants predict different
additional (s)particles, some of which (being colored) could give rise
to spectacular resonances at the LHC.  However, even if the new gauge
symmetry and all additional fields are outside the reach of the LHC,
all variants have different $\beta$ coefficients and thus different
running of MSSM parameters, both gauge couplings and SUSY soft
masses. Thus, if one assumes the validity of a certain SUSY breaking
scheme, such as for example mSugra, indirect traces of the different
variants remain in the SUSY spectrum, potentially measurable at the
LHC and a future ILC/CLIC. This was discussed earlier in the context
of indirect tests for the SUSY seesaw mechanism in
\cite{Buckley:2006nv,Hirsch:2008gh,Esteves:2010ff} and for extended
gauge models in \cite{DeRomeri:2011ie}. We generalize the discussion
of \cite{DeRomeri:2011ie} and show how the ``invariants'',
i.e. certain combinations of SUSY soft breaking parameters, can
themselves be organized into a few classes, which in principle allow
to distinguish class-II models from class-I or class-III and, if
sufficient precision could be reached experimentally, even select
specific variants within a class and give indirect information about
the new energy scale(s).

The rest of this paper is organized as follows: in the next section 
we first lay out the general conditions for the construction of the models 
we are interested in, before discussing variants and example configurations 
for all of the three classes we consider. Section \ref{sect:invariants}
then discusses ``invariants'', i.e. SUSY soft parameters in the 
different model classes. We then close with a short summary and 
discussion. Several technical aspects of our work are presented 
in the appendix.

\section{Models}
\label{sec:models}

\subsection{Supersymmetric SO(10) models: General considerations}
\label{subsec:so10models}

Before entering into the details of the different model classes, we
will first list some general requirements which we use in all
constructions.  These requirements are the basic conditions any 
model has to fulfill to guarantee at least in principle that a
phenomenologically realistic model will result.

We use the following conditions:
\begin{itemize}
\item Perturbative SO(10) unification. That is, gauge couplings unify 
(at least) as well as in the MSSM and the value of $\alpha_{G}$ is 
in the perturbative regime.
\item The GUT scale should lie above (roughly) $10^{16}$ GeV. 
This bound is motivated by the limit on the proton decay half-live. 
\item Sliding mechanism. This requirement is a set of conditions 
(different conditions for different classes of models) on the 
allowed $\beta$ coefficients of the gauge couplings, which ensure 
the additional gauge group structure can be broken at any energy 
scale consistent with GCU.
\item Renormalizable symmetry breaking. This implies that at each 
intermediate step we assume there are (at least) the minimal number 
of Higgs fields, which the corresponding symmetry breaking scheme 
requires.
\item Fermion masses and in particular neutrino masses. This condition
implies that the field content of the extended gauge groups is rich
enough to fit experimental data, although we will not attempt detailed
fits of all data. In particular, we require the fields 
to generate Majorana neutrino masses through seesaw, either ordinary
seesaw or inverse/linear seesaw, to be present.
\item Anomaly cancellation. We accept as valid ``models'' only field
configurations which are anomaly free.
\item $SO(10)$ completable. All fields used in a lower energy stage 
must be parts of a multiplet present at the next higher symmetry stage. 
In particular, all fields should come from the decomposition of one of 
the $SO(10)$ multiplets we consider (multiplets up to ${\bf 126}$).
\item Correct MSSM limit. All models must be rich enough in particle
content that at low energies the MSSM can emerge.
\end{itemize}

A few more words on our naming convention and notations might be
necessary. We consider the three different $SO(10)$ breaking chains,
eq. (\ref{eq:chainI})-(\ref{eq:chainIII}), and will call these model
``classes''. In each class there are fixed sets of
$\beta$-coefficients, which all lead to GCU but with different values
of $\alpha_{G}$ and different values of $\alpha_{R}$ and
$\alpha_{B-L}$ at low energies. These different sets are called
``variants'' in the following. And finally, (nearly) all of the
variants can be created by more than one possible set of superfields.
We will call such a set of superfields a ``configuration''.
Configurations are what usually is called ``model'' by model builders,
although we prefer to think of these as ``proto-models'',
i.e. constructions fulfilling all our basic requirements. These 
are only proto-models (and not full-fledged models), since we do 
not check for each configuration in a detailed calculation that all the 
fields required in that configuration can remain light. We believe 
that for many, but probably not all, of the configurations one can find 
conditions for the required field combinations being ``light'', following 
similar conditions as discussed in the prototype class-I model 
of \cite{Dev:2009aw}.

All superfields are named as $\Phi_{3_c,2_L,2_R,1_{B-L}}$ (in the left-right 
symmetric stage),  $\Psi_{4,2_L,2_R}$ (in the Pati-Salam regime) and 
$\Phi^{'}_{3_c,2_L,1_R,1_{B-L}}$ (in the $U(1)_R\times U(1)_{B-L}$ 
regime), with the indices giving the transformation properties under 
the group. A conjugate of a field is denoted by, for example,  
$\bar{\Phi}_{3_c,2_L,2_R,1_{B-L}}$, however, without putting a corresponding 
``bar'' (or minus sign) in the index. 
We list all fields we use, together with their transformation 
properties and their origin from $SO(10)$ multiplets, complete up to 
the ${\bf 126}$ of $SO(10)$ in the appendix.
 
\subsection{Model class-I: One intermediate (left-right) scale}
\label{subsect:LRm}

We start our discussion with the simplest class of models with only 
one new intermediate scale (LR):
\begin{equation}
SO(10)\rightarrow SU(3)_{c}\times SU(2)_{L} \times SU(2)_{R} \times
U(1)_{B-L} \rightarrow \text{MSSM} \ . 
\end{equation}
We do not discuss the first symmetry breaking step in detail, since 
it is not relevant for the following discussion and only mention that 
$SO(10)$ can be broken to the LR group either via the interplay of 
vevs from a ${\bf 45}$ and a ${\bf 54}$, as done for example in 
\cite{Dev:2009aw}, or via a  ${\bf 45}$ and a ${\bf 210}$, an approach 
followed in \cite{Malinsky:2005bi}. In the left-right symmetric 
stage we consider all irreducible representations, which can be 
constructed from $SO(10)$ multiplets up to dimension ${\bf 126}$. 
This allows for a total of 24 different representations (plus conjugates), 
their transformation properties under the LR group and their $SO(10)$ 
origin are summarized in table~\ref{tab:List_of_LR_fields} 
(and table~\ref{tab:List_of_PatiSalam_fields}) of the appendix.

Consider gauge coupling unification first. If we take the MSSM 
particle content as a starting point, the $\beta$-coefficients in the 
different regimes are given as: \footnote{For $b_{1}^{SM}$ and $b_{2}^{SM}$ 
we use the SM particle content plus one additional Higgs doublet.}
\begin{eqnarray}
\left(b_{3}^{SM},b_{2}^{SM},b_{1}^{SM}\right) & = & \left(-7,-3,21/5\right)\nonumber \\
\left(b_{3}^{MSSM},b_{2}^{MSSM},b_{1}^{MSSM}\right) & = & \left(-3,1,33/5\right)\nonumber \\
\left(b_{3}^{LR},b_{2}^{LR},b_{R}^{LR},b_{B-L}^{LR}\right) & = & \left(-3,1,1,6\right)+\left(\Delta b_{3}^{LR},\Delta b_{2}^{LR},\Delta b_{R}^{LR},\Delta b_{B-L}^{LR}\right)
\end{eqnarray}
where we have used the canonical normalization for $(B-L)$ related to 
the physical one by $(B-L)^{c}=\sqrt{\frac{3}{8}}(B-L)^{p}$. Here, 
$\Delta b_{i}^{LR}$ stands for the contributions from additional 
superfields, not accounted for in the MSSM.

\begin{figure}[htb]
\centering
\begin{tabular}{cc}
\includegraphics[width=0.6\linewidth]{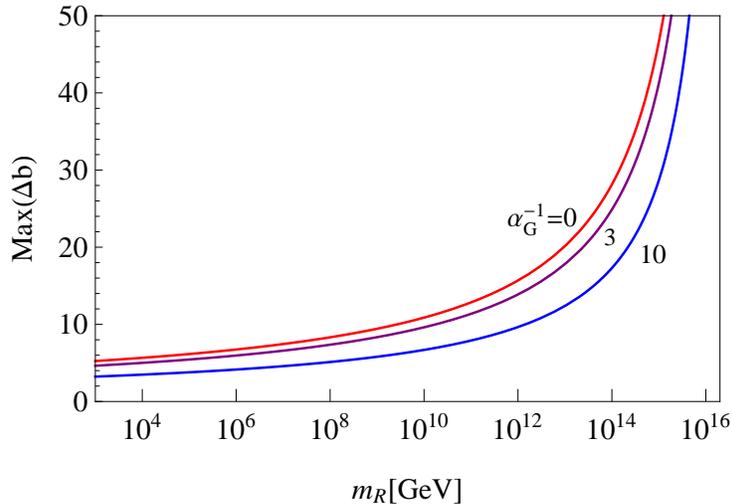}
\end{tabular}
\caption{Maximum value of $\Delta b$ allowed by perturbativity 
as function of the scale $m_{R}$ in GeV. The three different lines 
have been calculated for three different values for the unified 
coupling $\alpha_G^{-1}$, namely $\alpha_G^{-1}=0,3,10$. An LR 
scale below 10 TeV (1 TeV) requires Max$(\Delta b_3)\lsim 5.7$ 
($5.2$) if the extreme value of $\alpha_G^{-1}=0$ is chosen and 
Max$(\Delta b_3)\lsim 5.1$ ($4.7$) for $\alpha_G^{-1}=3$.}
\label{fig:MaxDb}  
\end{figure}

As is well known, while the MSSM unifies, putting an additional LR
scale below the GUT scale with $\forall\Delta b_{i}^{LR}=0$ destroys
unification. Nevertheless GCU can be maintained, if some simple conditions
on the $\Delta b_{i}^{LR}$ are fulfilled. First, since in the MSSM
$\alpha_{3}=\alpha_{2}$ at roughly $2\times10^{16}$ GeV one has
that $\Delta b_{2}^{LR}=\Delta b_{3}^{LR}\equiv\Delta b$ in order
to preserve this situation for an arbitrary LR scale (sliding condition).
Next, recall the matching condition 
\begin{equation}
\alpha_{1}^{-1}(m_{R})=\frac{3}{5}\alpha_{R}^{-1}(m_{R})+\frac{2}{5}\alpha_{B-L}^{-1}(m_{R})\label{eq:match}
\end{equation}
which, by substitution of the LR scale by an arbitrary one above
$m_{R}$, allows us to define an artificial continuation of the
hypercharge coupling constant $\alpha_{1}$ into the LR stage. The
$\beta$-coefficient of this dummy coupling constant for $E>m_{R}$ is
$\frac{3}{5}b_{R}^{LR}+\frac{2}{5}b_{B-L}^{LR}$ and it should be
compared with $b_{1}^{MSSM}$ ($E<m_{R}$); the difference is
$\frac{3}{5}\Delta b_{R}^{LR}+\frac{2}{5}\Delta
b_{B-L}^{LR}-\frac{18}{5}$ and it must be equal to $\Delta b$ in order
for the difference between this $\alpha_{1}$ coupling and
$\alpha_{3}=\alpha_{2}$ at the GUT to be independent of the scale
$m_{R}$. These are the two conditions imposed by the sliding
requirement of the LR scale on the $\beta$-coefficients 
[see eq. (\ref{eq:variantsLR})]. Note, however, that we did not require
(approximate) unification of $\alpha_{R}$ and $\alpha_{B-L}$ with
$\alpha_{3}$ and $\alpha_{2}$; it was sufficient to require that
$\alpha_{2}^{-1}=
\alpha_{3}^{-1}\approx\frac{3}{5}\alpha_{R}^{-1}+\frac{2}{5}\alpha_{B-L}^{-1}$.
In any case, we can always achieve the desired unification because the
splitting between $\alpha_{R}$ and $\alpha_{B-L}$ at the $m_{R}$ scale
is a free parameter, so it can be used to force
$\alpha_{R}=\alpha_{B-L}$ at the scale where $\alpha_{3}$ and
$\alpha_{2}$ unify, which leads to an almost perfect unification of
the four couplings. Also, we require that unification is perturbative,
i.e. the value of the common coupling constant at the GUT scale is
$\alpha_G^{-1}\ge 0$.  From the experimental value of
$\alpha_S(m_{Z})$ \cite{Beringer:1900zz} one can easily calculate
the maximal allowed value of $\Delta b$ as a function of the scale,
where the LR group is broken to the SM group. This is shown in
fig.~\ref{fig:MaxDb} for three different values of
$\alpha_G^{-1}$. The smallest Max($\Delta b$) is obtained for the
smallest value of $m_R$ (and the largest value of $\alpha_G^{-1}$).
For $\alpha_G^{-1}$ in the interval $[0,3]$ one obtains Max($\Delta
b$) in the range [$4.7,5.7$], i.e. we will study cases up to a
Max$(\Delta b)=5$ (see, however, the discussion below).

\begin{figure}[htb]
\centering
\begin{tabular}{cc}
\includegraphics[width=0.45\linewidth]{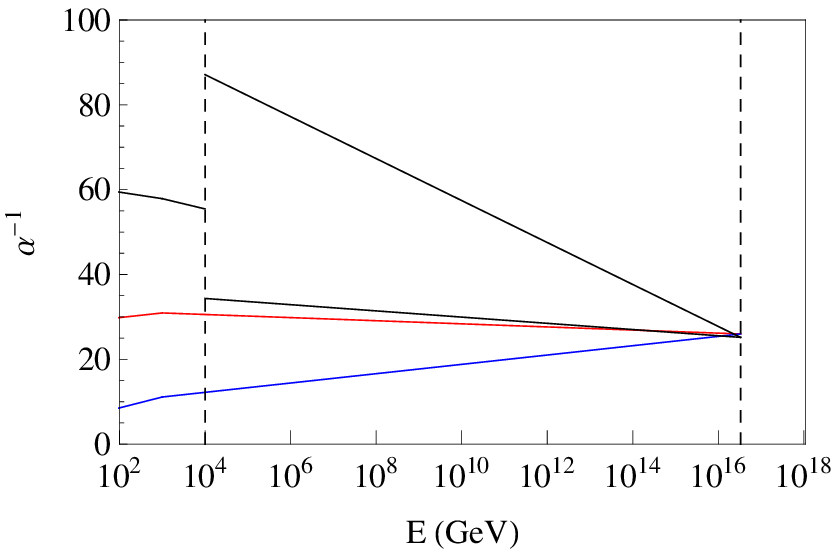}&  
\includegraphics[width=0.45\linewidth]{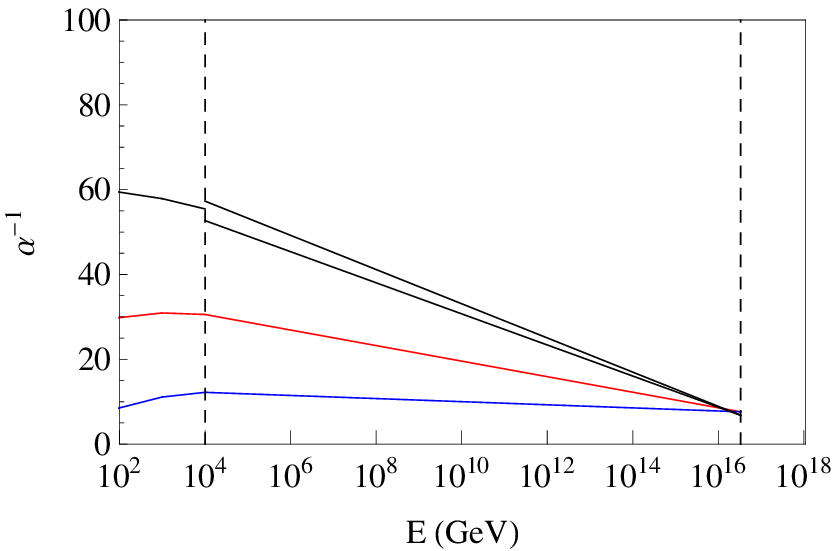}
\end{tabular}
\caption{Gauge coupling unification in LR models for $m_{R}=10^{4}$
  GeV. Left panel is for ($\Delta b^{LR}_{3},\Delta b^{LR}_{2},\Delta
  b^{LR}_{R},\Delta b^{LR}_{B-L}$) $=(0,0,1,15/2)$ and right panel for
  $(4,4,10,4)$.}
\label{fig:LR-Running}  
\end{figure}

All together these considerations result in the following constraints 
on the allowed values for the $\Delta b^{LR}_i$:
\begin{eqnarray}\label{eq:variantsLR}
\Delta b^{LR}_{2} = \Delta b^{LR}_{3} =\Delta b &\le & 5 ,\\[+2mm]\nonumber
\Delta b_{B-L}^{LR} +\frac{3}{2}\Delta b^{LR}_{R} -9 
 = \frac{5}{2}\Delta b &\le &\frac{25}{2} .
\end{eqnarray}
Given eq. (\ref{eq:variantsLR}) one can calculate all allowed variants 
of sets of $\Delta b^{LR}_i$, guaranteed to give GCU. Two examples 
are shown in fig. ~\ref{fig:LR-Running}. The figure shows the running 
of the inverse gauge couplings as a function of the energy scale, 
for an assumed value of $m_R=10$ TeV and a SUSY scale of 1 TeV, for 
($\Delta b^{LR}_{3},\Delta b^{LR}_{2},\Delta b^{LR}_{R},\Delta b^{LR}_{B-L}$)
$=(0,0,1,15/2)$ (left) and $=(4,4,10,4)$ (right). The example on the left 
has $\alpha_G^{-1} \simeq 25$ as in the MSSM, while the example on the 
right has $\alpha_G^{-1} \simeq 6$. Note that while both examples 
lead by construction to the same value of $\alpha_1(m_Z)$, they 
have very different values for $\alpha_{R}(m_R)$ and  $\alpha_{B-L}(m_R)$ 
and thus predict different couplings for the gauge bosons $W_R$ and 
$Z'$ of the extended gauge group.

With the constraints from eq. (\ref{eq:variantsLR}), we find that a
total of 65 different variants can be constructed. However, after
imposing that at least one of the fields that breaks correctly the
$SU(2)_{R} \times U(1)_{B-L}$ symmetry to $U(1)_Y$ is present, either
a $\Phi_{1,1,3,-2}$ or a $\Phi_{1,1,2,-1}$ (and/or their conjugates),
the number of variants is reduced to 53. We list them in
tables~\ref{tab:LR_field_configuration_A} and
\ref{tab:LR_field_configuration_B}, together with one example of field
configurations which give the corresponding $\Delta b^{LR}_i$.

\begin{table}
\begin{center}
\scalebox{0.95}{
\begin{tabular}{ll}
\toprule 
$(\Delta b,\Delta b_{R})$ & Sample field combination \tabularnewline
\midrule 
(0, 1) & $\bar{\Phi}_{1,1,2,-1}+2\bar{\Phi}_{1,1,1,2}+\Phi_{1,1,2,-1}+2\Phi_{1,1,1,2}$\tabularnewline
(0, 2) & $2\bar{\Phi}_{1,1,2,-1}+\bar{\Phi}_{1,1,1,2}+2\Phi_{1,1,2,-1}+\Phi_{1,1,1,2}$\tabularnewline
(0, 3) & $\bar{\Phi}_{1,1,2,-1}+\bar{\Phi}_{1,1,1,2}+\Phi_{1,1,2,-1}+\Phi_{1,1,3,0}+\Phi_{1,1,1,2}$\tabularnewline
(0, 4) & $2\bar{\Phi}_{1,1,2,-1}+2\Phi_{1,1,2,-1}+\Phi_{1,1,3,0}$\tabularnewline
(0, 5) & $\bar{\Phi}_{1,1,2,-1}+\Phi_{1,1,2,-1}+2\Phi_{1,1,3,0}$\tabularnewline
(1, 1) &
$\bar{\Phi}_{1,2,1,1}+\bar{\Phi}_{1,1,2,-1}+2\bar{\Phi}_{1,1,1,2}+\bar{\Phi}_{3,1,1,-\frac{2}{3}}+\Phi_{1,2,1,1}+\Phi_{1,1,2,-1}+2\Phi_{1,1,1,2}+\Phi_{3,1,1,-\frac{2}{3}}$\tabularnewline
 (1, 2) &
 $\bar{\Phi}_{1,1,2,-1}+2\bar{\Phi}_{1,1,1,2}+\bar{\Phi}_{3,1,1,-\frac{2}{3}}+\Phi_{1,1,2,-1}+\Phi_{1,2,2,0}+2\Phi_{1,1,1,2}+\Phi_{3,1,1,-\frac{2}{3}}$\tabularnewline
 (1, 3) &
 $2\bar{\Phi}_{1,1,2,-1}+\bar{\Phi}_{1,1,1,2}+\bar{\Phi}_{3,1,1,-\frac{2}{3}}+2\Phi_{1,1,2,-1}+\Phi_{1,2,2,0}+\Phi_{1,1,1,2}+\Phi_{3,1,1,-\frac{2}{3}}$\tabularnewline
 (1, 4) &
 $\bar{\Phi}_{1,1,2,-1}+\bar{\Phi}_{1,1,1,2}+\bar{\Phi}_{3,1,1,-\frac{2}{3}}+\Phi_{1,1,2,-1}+\Phi_{1,1,3,0}+\Phi_{1,2,2,0}+\Phi_{1,1,1,2}+\Phi_{3,1,1,-\frac{2}{3}}$\tabularnewline
 (1, 5) & $2\bar{\Phi}_{1,1,2,-1}+\bar{\Phi}_{3,1,1,-\frac{2}{3}}+2\Phi_{1,1,2,-1}+\Phi_{1,1,3,0}+\Phi_{1,2,2,0}+\Phi_{3,1,1,-\frac{2}{3}}$\tabularnewline
 (1, 6) & $\bar{\Phi}_{1,1,2,-1}+\bar{\Phi}_{3,1,1,-\frac{2}{3}}+\Phi_{1,1,2,-1}+2\Phi_{1,1,3,0}+\Phi_{1,2,2,0}+\Phi_{3,1,1,-\frac{2}{3}}$\tabularnewline
 (2, 1) &
 $\bar{\Phi}_{1,1,2,-1}+3\bar{\Phi}_{1,1,1,2}+2\bar{\Phi}_{3,1,1,-\frac{2}{3}}+\Phi_{1,1,2,-1}+\Phi_{1,3,1,0}+3\Phi_{1,1,1,2}+2\Phi_{3,1,1,-\frac{2}{3}}$\tabularnewline
 (2, 2) &
 $2\bar{\Phi}_{1,1,2,-1}+2\bar{\Phi}_{1,1,1,2}+2\bar{\Phi}_{3,1,1,-\frac{2}{3}}+2\Phi_{1,1,2,-1}+\Phi_{1,3,1,0}+2\Phi_{1,1,1,2}+2\Phi_{3,1,1,-\frac{2}{3}}$\tabularnewline
 (2, 3) &
 $\bar{\Phi}_{1,1,2,-1}+2\bar{\Phi}_{1,1,1,2}+2\bar{\Phi}_{3,1,1,-\frac{2}{3}}+\Phi_{1,1,2,-1}+2\Phi_{1,2,2,0}+2\Phi_{1,1,1,2}+2\Phi_{3,1,1,-\frac{2}{3}}$\tabularnewline
 (2, 4) &
 $2\bar{\Phi}_{1,1,2,-1}+\bar{\Phi}_{1,1,1,2}+2\bar{\Phi}_{3,1,1,-\frac{2}{3}}+2\Phi_{1,1,2,-1}+2\Phi_{1,2,2,0}+\Phi_{1,1,1,2}+2\Phi_{3,1,1,-\frac{2}{3}}$\tabularnewline 
 (2, 5) & $\bar{\Phi}_{1,1,2,-1}+\bar{\Phi}_{1,1,1-2}+2\bar{\Phi}_{3,1,1,-\frac{2}{3}}+\Phi_{1,1,2,-1}+\Phi_{1,1,3,0}+2\Phi_{1,2,2,0}+\Phi_{1,1,1,2}+2\Phi_{3,1,1,-\frac{2}{3}}$\tabularnewline
 (2, 6) &
 $2\bar{\Phi}_{1,1,2,-1}+2\bar{\Phi}_{3,1,1,-\frac{2}{3}}+2\Phi_{1,1,2,-1}+\Phi_{1,1,3,0}+2\Phi_{1,2,2,0}+2\Phi_{3,1,1,-\frac{2}{3}}$\tabularnewline
 (2, 7) & $\bar{\Phi}_{1,1,2,-1}+2\bar{\Phi}_{3,1,1,-\frac{2}{3}}+\Phi_{1,1,2,-1}+2\Phi_{1,1,3,0}+2\Phi_{1,2,2,0}+2\Phi_{3,1,1,-\frac{2}{3}}$\tabularnewline
 (2, 8) & $\bar{\Phi}_{1,1,2,-1}+\bar{\Phi}_{3,1,2,\frac{1}{3}}+\Phi_{1,1,2,-1}+\Phi_{1,1,3,0}+2\Phi_{1,2,2,0}+\Phi_{3,1,2,\frac{1}{3}}$\tabularnewline
 (3, 1) & $\bar{\Phi}_{1,2,1,1} +\bar{\Phi}_{1,1,2,-1}+4\bar{\Phi}_{1,1,1,2}+\Phi_{1,2,1,1}+\Phi_{1,1,2,-1}+\Phi_{1,3,1,0}+\Phi_{8,1,1,0}+4\Phi_{1,1,1,2}$\tabularnewline
 (3, 2) &
 $\bar{\Phi}_{1,1,2,-1}+ 4\bar{\Phi}_{1,1,1,2}+\Phi_{1,1,2,-1}+\Phi_{1,3,1,0}+\Phi_{1,2,2,0}+\Phi_{8,1,1,0}+4\Phi_{1,1,1,2}$ \tabularnewline
 (3, 3) & $2\bar{\Phi}_{1,1,2,-1}+3\bar{\Phi}_{1,1,1,2}+2\Phi_{1,1,2,-1}+\Phi_{1,3,1,0}+\Phi_{1,2,2,0}+\Phi_{8,1,1,0}+3\Phi_{1,1,1,2}$\tabularnewline
 (3, 4) & $\bar{\Phi}_{1,2,1,1}+\bar{\Phi}_{1,1,3,-2}+\Phi_{1,2,1,1}+\Phi_{1,3,1,0}+\Phi_{8,1,1,0}+\Phi_{1,1,3,-2}$\tabularnewline
 (3, 5) & $\bar{\Phi}_{1,1,3,-2}+\Phi_{1,3,1,0}+\Phi_{1,2,2,0}+\Phi_{8,1,1,0}+\Phi_{1,1,3,-2}$\tabularnewline
 (3, 6) &
 $\bar{\Phi}_{1,1,2,-1}+2\bar{\Phi}_{1,1,1,2}+\Phi_{1,1,2,-1}+\Phi_{1,1,3,0}+3\Phi_{1,2,2,0}+\Phi_{8,1,1,0}+2\Phi_{1,1,1,2}$ \tabularnewline
 (3, 7) & $2\bar{\Phi}_{1,1,2,-1}+\bar{\Phi}_{1,1,1-2}+2\Phi_{1,1,2,-1}+\Phi_{1,1,3,0}+3\Phi_{1,2,2,0}+\Phi_{8,1,1,0}+\Phi_{1,1,1,2}$\tabularnewline
 (3, 8) &
 $\bar{\Phi}_{1,1,2,-1}+\bar{\Phi}_{1,1,1,2}+\Phi_{1,1,2,-1}+2\Phi_{1,1,3,0}+3\Phi_{1,2,2,0}+\Phi_{8,1,1,0}+\Phi_{1,1,1,2}$ \tabularnewline
 (3, 9) & $2\bar{\Phi}_{1,1,2,-1}+2\Phi_{1,1,2,-1}+2\Phi_{1,1,3,0}+3\Phi_{1,2,2,0}+\Phi_{8,1,1,0}$\tabularnewline
 (3, 10) & $\bar{\Phi}_{1,1,2,-1}+\Phi_{1,1,2,-1}+3\Phi_{1,1,3,0}+3\Phi_{1,2,2,0}+\Phi_{8,1,1,0}$\tabularnewline
\bottomrule
\end{tabular}
}
\end{center}
\caption{List of the 53 variants with a single LR scale. Shown are the
29 variants with $\Delta b_{3}<4$. In each case, the fields shown are
the extra ones which are needed besides the ones contained in the MSSM
representations (the 2 Higgs doublets are assumed to come from one
bi-doublet $\Phi_{1,2,2,0}$).  The $\Delta b_{3},\Delta b_{2},\Delta
b_{R},\Delta b_{B-L}$ values can be obtained from the first column
through eqs~(\ref{eq:variantsLR}).}
\label{tab:LR_field_configuration_A}
\end{table}

\begin{table}
\begin{center}
\scalebox{0.92}{
\begin{tabular}{ll}
\toprule 
$(\Delta b,\Delta b_{R})$ & Sample field combination \tabularnewline
\midrule 
 (4, 1) &
 $\bar{\Phi}_{1,1,2,-1}+5\bar{\Phi}_{1,1,1,2}+\bar{\Phi}_{3,1,1,-\frac{2}{3}}+\Phi_{1,1,2,-1}+
 2\Phi_{1,3,1,0}+\Phi_{8,1,1,0}+5\Phi_{1,1,1,2}+\Phi_{3,1,1,-\frac{2}{3}}$\tabularnewline   
 (4, 2) &
 $2\bar{\Phi}_{1,1,2,-1}+4\bar{\Phi}_{1,1,1,2}+\bar{\Phi}_{3,1,1,-\frac{2}{3}}+2\Phi_{1,1,2,-1}+
 2\Phi_{1,3,1,0}+\Phi_{8,1,1,0}+4\Phi_{1,1,1,2}+\Phi_{3,1,1,-\frac{2}{3}}$ \tabularnewline 
 (4, 3) &
 $\bar{\Phi}_{1,1,2,-1}+4\bar{\Phi}_{1,1,1,2}+\bar{\Phi}_{3,1,1,-\frac{2}{3}}+\Phi_{1,1,2,-1}+\Phi_{1,3,1,0}+2\Phi_{1,2,2,0}+\Phi_{8,1,1,0}+4\Phi_{1,1,1,2}+\Phi_{3,1,1,-\frac{2}{3}}$\tabularnewline 
 (4, 4) &
 $\bar{\Phi}_{1,1,1,2}+\bar{\Phi}_{3,1,1,-\frac{2}{3}}+\bar{\Phi}_{1,1,3,-2}+
 2\Phi_{1,3,1,0}+\Phi_{8,1,1,0}+\Phi_{1,1,1,2}+\Phi_{3,1,1,-\frac{2}{3}}+\Phi_{1,1,3,-2}$ \tabularnewline   
 (4, 5) & $\bar{\Phi}_{1,1,2,-1}+\bar{\Phi}_{3,1,1,-\frac{2}{3}}+\bar{\Phi}_{1,1,3,-2}+\Phi_{1,1,2,-1}+2\Phi_{1,3,1,0}+\Phi_{8,1,1,0}+\Phi_{3,1,1,-\frac{2}{3}}+\Phi_{1,1,3,-2}$\tabularnewline
 (4, 6) &
 $\bar{\Phi}_{3,1,1,-\frac{2}{3}}+\bar{\Phi}_{1,1,3,-2}+\Phi_{1,3,1,0}+2\Phi_{1,2,2,0}+\Phi_{8,1,1,0}+\Phi_{3,1,1,-\frac{2}{3}}+\Phi_{1,1,3,-2}$\tabularnewline  
 (4, 7) & $\bar{\Phi}_{1,1,2,-1}+2\bar{\Phi}_{1,1,1,2}+\bar{\Phi}_{3,1,1,-\frac{2}{3}}+\Phi_{1,1,2,-1}+\Phi_{1,1,3,0}+4\Phi_{1,2,2,0}+\Phi_{8,1,1,0}+2\Phi_{1,1,1,2}+\Phi_{3,1,1,-\frac{2}{3}}$\tabularnewline
 (4, 8) & $2\bar{\Phi}_{1,1,2,-1}+\bar{\Phi}_{1,1,1,2}+\bar{\Phi}_{3,1,1,-\frac{2}{3}}+2\Phi_{1,1,2,-1}+\Phi_{1,1,3,0}+4\Phi_{1,2,2,0}+\Phi_{8,1,1,0}+\Phi_{1,1,1,2}+\Phi_{3,1,1,-\frac{2}{3}}$\tabularnewline 
 (4, 9) & $\bar{\Phi}_{1,1,2,-1}+\bar{\Phi}_{1,1,1,2}+\bar{\Phi}_{3,1,1,-\frac{2}{3}}+\Phi_{1,1,2,-1}+2\Phi_{1,1,3,0}+4\Phi_{1,2,2,0}+\Phi_{8,1,1,0}+\Phi_{1,1,1,2}+\Phi_{3,1,1,-\frac{2}{3}}$\tabularnewline
 (4, 10) & $2\bar{\Phi}_{1,1,2,-1}+\bar{\Phi}_{3,1,1,-\frac{2}{3}}+2\Phi_{1,1,2,-1}+2\Phi_{1,1,3,0}+4\Phi_{1,2,2,0}+\Phi_{8,1,1,0}+\Phi_{3,1,1,-\frac{2}{3}}$\tabularnewline
 (4, 11) & $\bar{\Phi}_{1,1,2,-1}+\bar{\Phi}_{3,1,1,-\frac{2}{3}}+\Phi_{1,1,2,-1}+3\Phi_{1,1,3,0}+4\Phi_{1,2,2,0}+\Phi_{8,1,1,0}+\Phi_{3,1,1,-\frac{2}{3}}$\tabularnewline
 (5, 1) & $\bar{\Phi}_{1,2,1,1}+\bar{\Phi}_{1,1,2,-1}+5\bar{\Phi}_{1,1,1,2}+2\bar{\Phi}_{3,1,1,-\frac{2}{3}}+\Phi_{1,2,1,1}+\Phi_{1,1,2,-1}+2\Phi_{1,3,1,0}+\Phi_{8,1,1,0}$\tabularnewline
  & $+5\Phi_{1,1,1,2}+2\Phi_{3,1,1,-\frac{2}{3}}$\tabularnewline
 (5, 2) & $\bar{\Phi}_{1,1,2,-1}+5\bar{\Phi}_{1,1,1,2}+2\bar{\Phi}_{3,1,1,-\frac{2}{3}}+\Phi_{1,1,2,-1}+2\Phi_{1,3,1,0}+\Phi_{1,2,2,0}+\Phi_{8,1,1,0}+5\Phi_{1,1,1,2}$\tabularnewline
 & $+2\Phi_{3,1,1,-\frac{2}{3}}$\tabularnewline
 (5, 3) & $2\bar{\Phi}_{1,1,2,-1}+4\bar{\Phi}_{1,1,1,2}+2\bar{\Phi}_{3,1,1,-\frac{2}{3}}+2\Phi_{1,1,2,-1}+2\Phi_{1,3,1,0}+\Phi_{1,2,2,0}+\Phi_{8,1,1,0}+4\Phi_{1,1,1,2}$\tabularnewline
  & $+2\Phi_{3,1,1,-\frac{2}{3}}$\tabularnewline
 (5, 4) & $\bar{\Phi}_{1,2,1,1}+\bar{\Phi}_{1,1,1,2}+2\bar{\Phi}_{3,1,1,-\frac{2}{3}}+\bar{\Phi}_{1,1,3,-2}+\Phi_{1,2,1,1}+2\Phi_{1,3,1,0}+\Phi_{8,1,1,0}+\Phi_{1,1,1,2}+2\Phi_{3,1,1,-\frac{2}{3}}$\tabularnewline
  & $+\Phi_{1,1,3,-2}$\tabularnewline
 (5, 5) & $\bar{\Phi}_{1,1,1,2}+2\bar{\Phi}_{3,1,1,-\frac{2}{3}}+\bar{\Phi}_{1,1,3,-2}+2\Phi_{1,3,1,0}+\Phi_{1,2,2,0}+\Phi_{8,1,1,0}+\Phi_{1,1,1,2}+2\Phi_{3,1,1,-\frac{2}{3}}$\tabularnewline
  & $+\Phi_{1,1,3,-2}$\tabularnewline
 (5, 6) & $\bar{\Phi}_{1,1,2,-1}+2\bar{\Phi}_{3,1,1,-\frac{2}{3}}+\bar{\Phi}_{1,1,3,-2}+\Phi_{1,1,2,-1}+2\Phi_{1,3,1,0}+\Phi_{1,2,2,0}+\Phi_{8,1,1,0}+2\Phi_{3,1,1,-\frac{2}{3}}$\tabularnewline
  & $+\Phi_{1,1,3,-2}$\tabularnewline
 (5, 7) & $2\bar{\Phi}_{3,1,1,-\frac{2}{3}}+\bar{\Phi}_{1,1,3,-2}+\Phi_{1,3,1,0}+3\Phi_{1,2,2,0}+\Phi_{8,1,1,0}+2\Phi_{3,1,1,-\frac{2}{3}}+\Phi_{1,1,3,-2}$\tabularnewline
 (5, 8) & $\bar{\Phi}_{3,1,2,\frac{1}{3}}+\bar{\Phi}_{1,1,3,-2}+2\Phi_{1,3,1,0}+\Phi_{1,2,2,0}+\Phi_{8,1,1,0}+\Phi_{3,1,2,\frac{1}{3}}+\Phi_{1,1,3,-2}$\tabularnewline
 (5, 9) &
 $2\bar{\Phi}_{1,1,2,-1}+\bar{\Phi}_{1,1,1,2}+2\bar{\Phi}_{3,1,1,-\frac{2}{3}}+2\Phi_{1,1,2,-1}+\Phi_{1,1,3,0}+5\Phi_{1,2,2,0}+\Phi_{8,1,1,0}+\Phi_{1,1,1,2}$ \tabularnewline
 &$+2\Phi_{3,1,1,-\frac{2}{3}}$ \tabularnewline
 (5, 10) & $\bar{\Phi}_{1,1,2,-1}+\bar{\Phi}_{1,1,1,2}+2\bar{\Phi}_{3,1,1,-\frac{2}{3}}+\Phi_{1,1,2,-1}+2\Phi_{1,1,3,0}+5\Phi_{1,2,2,0}+\Phi_{8,1,1,0}+\Phi_{1,1,1,2}$\tabularnewline
  & $+2\Phi_{3,1,1,-\frac{2}{3}}$\tabularnewline
 (5, 11) &
 $2\bar{\Phi}_{1,1,2,-1}+2\bar{\Phi}_{3,1,1,-\frac{2}{3}}+2\Phi_{1,1,2,-1}+2\Phi_{1,1,3,0}+5\Phi_{1,2,2,0}+\Phi_{8,1,1,0}+2\Phi_{3,1,1,-\frac{2}{3}}$
 \tabularnewline 
 (5, 12) & $\bar{\Phi}_{1,1,2,-1}+2\bar{\Phi}_{3,1,1,-\frac{2}{3}}+\Phi_{1,1,2,-1}+3\Phi_{1,1,3,0}+5\Phi_{1,2,2,0}+\Phi_{8,1,1,0}+2\Phi_{3,1,1,-\frac{2}{3}}$\tabularnewline
 (5, 13) &
 $\bar{\Phi}_{1,1,2,-1}+\bar{\Phi}_{3,1,2,\frac{1}{3}}+\Phi_{1,1,2,-1}+2\Phi_{1,1,3,0}+5\Phi_{1,2,2,0}+\Phi_{8,1,1,0}+\Phi_{3,1,2,\frac{1}{3}}$
 \tabularnewline
\bottomrule
\end{tabular}
}
\end{center}
\caption{List of the 53 variants with a single LR scale. Shown are the 
remaining 24 variants,  with $\Delta b_{3}\ge 4$.}
\label{tab:LR_field_configuration_B}
\end{table}

We give only one example for each configuration in
tables~\ref{tab:LR_field_configuration_A} and
\ref{tab:LR_field_configuration_B}, although we went through the 
exercise of finding all possible configurations for the 53 variants 
with the field content of table~\ref{tab:List_of_LR_fields}. In 
total there are 5324 anomaly-free configurations \cite{wp}. 
Only the variants (0,1), (0,2), (0,4) and (0,5) have only one configuration, 
while larger numbers of configurations are usually found for larger values of 
$\Delta b^{LR}_3$.

Not all the fields in table~\ref{tab:List_of_LR_fields} can lead to
valid configurations. The fields which never give an anomaly-free 
configuration are: $\Phi_{8,2,2,0}$, $\Phi_{3,2,2,\frac{4}{3}}$,
$\Phi_{3,3,1,-\frac{2}{3}}$, $\Phi_{3,1,3,-\frac{2}{3}}$,
$\Phi_{6,3,1,\frac{2}{3}}$, $\Phi_{6,1,3,\frac{2}{3}}$ and
$\Phi_{1,3,3,0}$. Also the field $\Phi_{3,2,2,-\frac{2}{3}}$ appears
only exactly once in the variant (5,5) in the configuration
$4\Phi_{1,2,1,1} + \Phi_{3,1,1,-\frac{2}{3}} +
\Phi_{3,2,2,-\frac{2}{3}} + 4 \Phi_{1,1,2,1} + 2 \Phi_{1,1,1,2} + 5
\Phi_{3,1,1,-\frac{2}{3}}$.  Note that, the example configurations we
give for the variants (1,3) and (1,4) are not the model-II and model-I
discussed in \cite{DeRomeri:2011ie}.

Many of the 53 variants have only configurations with
$\Phi_{1,1,2,-1}$ (and conjugate) for the breaking of the
LR-symmetry. These variants need either the presence of
$\Phi_{1,3,1,0}$ [as for example in the configuration shown for
  variant (2,1)] or $\Phi_{1,1,3,0}$ [see, for example (1,4)] or an
additional singlet $\Phi_{1,1,1,0}$ (not shown, since no contribution
to any $\Delta b^{LR}_i$), to generate seesaw neutrino masses. Using
the $\Phi_{1,1,1,0}$ one could construct either an inverse
\cite{mohapatra:1986bd} or a linear
\cite{Akhmedov:1995ip,Akhmedov:1995vm} seesaw mechanism, while with
$\Phi_{1,3,1,0}$ a seesaw type-III \cite{Foot:1988aq} is a possibility
and, finally a $\Phi_{1,1,3,0}$ allows for an inverse seesaw type-III
\cite{DeRomeri:2011ie}.  The first example where a valid configuration
with $\Phi_{1,1,3,-2}$ appears is the variant (3,4). The simplest
configuration is
$\Phi_{1,2,1,1}+\Phi_{1,3,1,0}+\Phi_{8,1,1,0}+\Phi_{1,1,3,-2}+ {\bar{\Phi}_{1,2,1,1}}
+{\bar{\Phi}_{1,1,3,-2}}$ (not the
example given in table~\ref{tab:LR_field_configuration_A}).  The vev
of the $\Phi_{1,1,3,-2}$ does not only break the LR symmetry, it can
also generate a Majorana mass term for the right-handed neutrino
fields, i.e. configurations with $\Phi_{1,1,3,-2}$ can generate a
seesaw type-I, in principle. Finally, the simplest possibility with a
valid configuration including $\Phi_{1,3,1,-2}$ is found in variant
(4,1) with
$\Phi_{1,1,2,-1}+\Phi_{8,1,1,0}+\Phi_{1,1,1,2}+\Phi_{3,1,1,\frac{4}{3}}+
\Phi_{1,3,1,-2}+\bar{\Phi}_{1,1,2,-1}+\bar{\Phi}_{1,1,1,2} +
\bar{\Phi}_{3,1,1,\frac{4}{3}}+\bar{\Phi}_{1,3,1,-2}$.  
The presence of $\Phi_{1,3,1,-2}$ allows to generate a seesaw type-II 
for the neutrinos.

As mentioned in the introduction, it is not possible to construct 
a sliding scale model in which the LR symmetry is broken by two 
pairs of triplets: $\Phi_{1,3,1,-2}+\bar{\Phi}_{1,3,1,-2}+
\Phi_{1,1,3,-2}+\bar{\Phi}_{1,1,3,-2}$. The sum of the $\Delta b$'s 
for these fields adds up to $(\Delta
b^{LR}_{3},b^{LR}_{L}, \Delta b^{LR}_{R}, \Delta b^{LR}_{B-L})=(0,4,4,18)$. 
This leaves only the possibilities ($4,4$), ($5,4$), ($5,5$), etc. from 
table \ref{tab:LR_field_configuration_B}. However, the largest 
$\Delta b^{LR}_{B-L}$ of these models is ($5,4$) which allows for 
$\Delta b^{LR}_{B-L}=31/2$, smaller than the required 18. This 
observation is consistent with the analysis done in 
\cite{Majee:2007uv}, where the authors have shown that a 
supersymmetric LR-symmetric model, where the LR symmetry is broken 
by two pairs of triplets, requires a minimal LR scale of at least 
$10^9$ GeV (and, actually, a much larger scale in minimal renormalizable 
models, if GUT scale thresholds are small).

A few final comments on the variants with $\Delta b^{LR}_2=\Delta b^{LR}_3=0$. 
Strictly speaking, none of these variants is guaranteed to give a 
valid model in the sense defined in sub-section \ref{subsec:so10models}, 
since they contain only one $\Phi_{1,2,2,0}\rightarrow (H_u,H_d)$ 
and no vector-like quarks (no $\Phi_{3,1,1,\frac{4}{3}}$ or 
$\Phi_{3,1,1,-\frac{2}{3}}$). With such a minimal configuration 
the CKM matrix is trivial at the energy scale where the LR 
symmetry is broken. We nevertheless list these variants, since 
in principle a CKM matrix for quarks consistent with experimental 
data could be generated at 1-loop level from flavor violating 
soft terms, as discussed in \cite{Babu:1998tm}. 

Before we end this section let us mention that variants with 
$\Delta b_3^{LR}=5$ will not be testable at LHC by measurements 
of soft SUSY breaking mass terms (``invariants''). This is 
discussed below in section \ref{sect:leadloginv}.

\subsection{Model class-II: Additional intermediate Pati-Salam scale}
\label{subsec:PSmodels}

In the second class of supersymmetric $SO(10)$ models we consider, 
$SO(10)$ is broken first to the Pati-Salam (PS) group. The complete breaking 
chain thus is:
\begin{eqnarray}
SO(10) &\to & SU(4) \times SU(2)_{L} \times SU(2)_{R} \\ \nonumber
   &\to & SU(3)_{c} \times SU(2)_{L} \times SU(2)_{R} \times
  U(1)_{B-L} \to \hskip2mm {\rm MSSM}\textrm{.}
\end{eqnarray}
The representations available from the decomposition of $SO(10)$ multiplets 
up to ${\bf 126}$ are listed in table~\ref{tab:List_of_PatiSalam_fields} 
in the appendix, together with their possible $SO(10)$ origin. 
Breaking  $SO(10)$ to the PS group requires that $\Psi_{1,1,1}$ from 
the ${\bf 54}$ takes a vev. The subsequent breaking of the PS group to 
the LR group requires that the singlet in $\Psi_{15,1,1}$, 
originally from the ${\bf 45}$ of $SO(10)$, acquires a vev. And, finally, 
as before in the LR-class, the breaking of LR to 
$SU(3)_{c} \times SU(2)_{L} \times U(1)_{Y}$ can be either done via 
$\Phi_{1,1,2,-1}$ or $\Phi_{1,1,3,-2}$ (and/or conjugates).

\begin{figure}[htb]
\centering
\begin{tabular}{cc}
\includegraphics[width=0.5\linewidth]{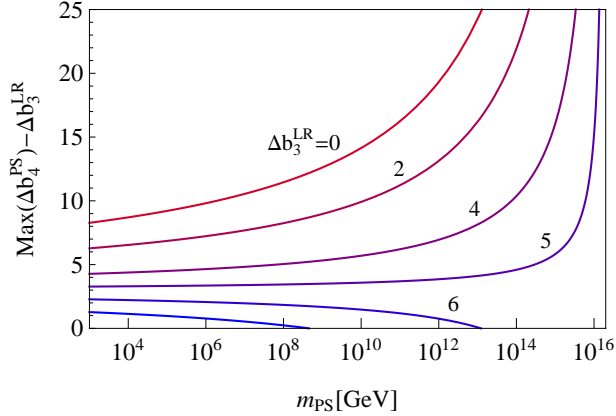}
\end{tabular}
\caption{Maximum value of $\Delta b^{PS}_4-\Delta b^{LR}_3$ allowed by
  perturbativity as function of the scale $m_{PS}$ in GeV. The
  different lines have been calculated for six different values of
  $\Delta b^{LR}_3$.  The plot assumes that $m_R = 1$ TeV. The line near the 
  bottom corresponds to $\Delta b_3^{LR}=7$.}
\label{fig:MaxDbPS}  
\end{figure}  
 
The additional $b_{i}$ coefficients for the regime $[m_{PS} ,
  m_{GUT}]$ are given by:
\begin{align}\label{eq:deltaBPS}
(b_{4}^{PS},b_{2}^{PS},b_{R}^{PS} ) &= ( -6,1,1) + ( \Delta
b_{4}^{PS},\Delta b_{2}^{PS},\Delta b_{R}^{PS} ) 
\end{align}
where, as before, the $\Delta b_i^{PS}$ include contributions from superfields 
not part of the MSSM field content. 

In this class of models, the unification scale is
independent of the LR one if the following condition is satisfied:
\begin{alignat}{1}
0= & \begin{pmatrix}\Delta b_{3}^{LR}-\Delta b_{2}^{LR} , & \frac{3}{5}\Delta b_{R}^{LR}+\frac{2}{5}\Delta b_{B-L}^{LR}-\Delta b_{2}^{LR}-\frac{18}{5}\end{pmatrix}.\begin{pmatrix}\begin{array}{rr}
2 & 3\\
-5 & 0
\end{array}\end{pmatrix}.\begin{pmatrix}\Delta b_{4}^{PS}-\Delta b_{2}^{PS}-3\\
\Delta b_{R}^{PS}-\Delta b_{2}^{PS}-12
\end{pmatrix}
\end{alignat}
It is worth noting that requiring also that $m_{PS}$ is independent of
the LR scale would lead to the conditions in
eq. (\ref{eq:variantsLR}), which are the sliding conditions for LR
models. We can see that this must be so in the following way: for some
starting values at $m_{PS}$ of the three gauge couplings, the scales
$m_{PS}$ and $m_{G}$ can be adjusted such that the two splittings
between the three gauge couplings are reduced to zero at $m_{G}$. This
fixes these scales, which must not change even if $m_{R}$ is
varied. As such
$\alpha_{3}^{-1}\left(m_{PS}\right)-\alpha_{2}^{-1}\left(m_{PS}\right)$
and
$\alpha_{3}^{-1}\left(m_{PS}\right)-\alpha_{R}^{-1}\left(m_{PS}\right)$
are also fixed and they can be determined by running the MSSM up to
$m_{PS}$. The situation is therefore equal to the one that lead to the
equalities in eq. (\ref{eq:variantsLR}), namely the splittings between the
gauge couplings at some fixed scale must be independent of $m_{R}$.

Since there are now two unknown scales involved in the problem, the
maximum $\Delta b^{X}_i$ allowed by perturbativity in one regime do
not only depend on the new scale $X$, but also on the $\Delta b^{Y}_i$
in the other regime as well. As an example, in fig. ~\ref{fig:MaxDbPS}
we show the Max($\Delta b^{PS}_4$) allowed by $\alpha_{G}^{-1}\ge 0$
for different values of $\Delta b^{LR}_3$ and for the choice $m_R = 1$
TeV and and $m_G = 10^{16}$ GeV.  The dependence of Max($\Delta
b^{PS}_4$) on $m_R$ is rather weak, as long as $m_R$ does not approach
the GUT scale.

\begin{figure}[htb]
\centering
\includegraphics[scale=0.26]{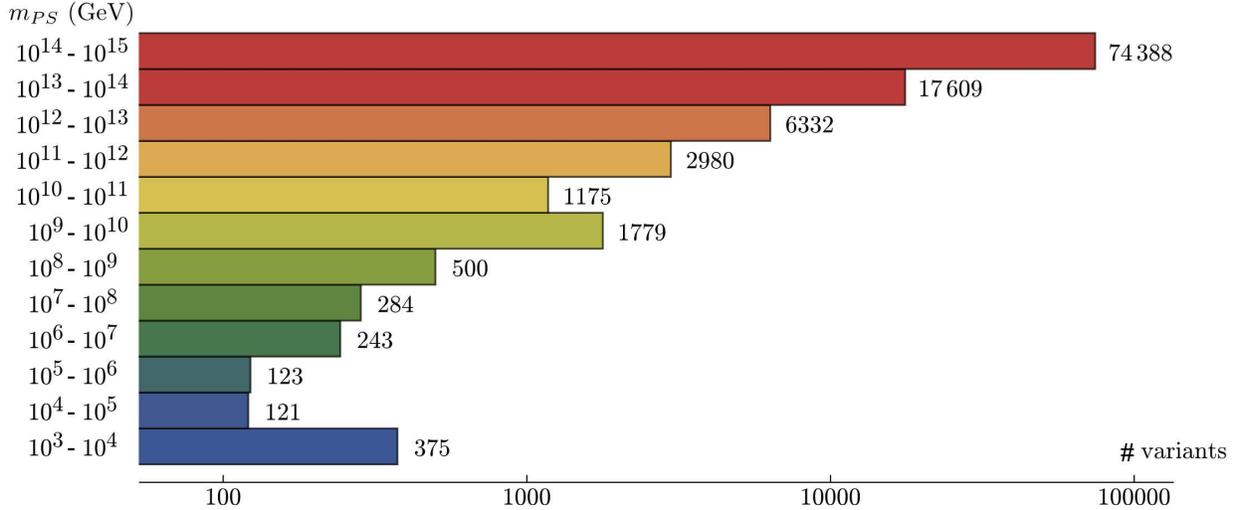}
\begin{tabular}{cc}
\end{tabular}
\caption{The number of possible variants in model class-II, 
assuming $m_R$ is of order $m_R\simeq 1$ TeV as a function of 
$m_{PS}$. Up to $m_{PS} = 10^{15}$ GeV the 
list is exhaustive. For larger values of $m_{PS}$ we have only 
scanned a finite (though large) set of possible variants.}
\label{fig:ScanConfigsPS}
\end{figure}

If we impose the limits $m_{R}=10^{3}$ GeV, $m_{PS} \leq 10^{6}$ GeV 
and take  $m_{G}=10^{16}$ GeV, the bounds for the different 
$\Delta b^{'}s$ can be written as:  %
\footnote{In fact, the bounds shown here exclude a few variants with $m_{PS}<10^{6}$
GeV. This is because of the following: while in most cases the most
conservative assumption is to assume that $m_{PS}$ is as large as
possible ($=10^{6}$ GeV; this leads to a smaller running in the PS
regime) in deriving these bounds, there are some cases where this
is not true. This is a minor complication which nonetheless was taken
into account in our computations.%
}
\begin{align}
\Delta b_{2}^{PS} +\frac{3}{10}\Delta b_{2}^{LR} &<7.2\\
\Delta b_{4}^{PS} +\frac{3}{10}\Delta b_{3}^{LR}&<10\\
\frac{2}{5}\Delta b_{4}^{PS}+\frac{3}{5}\Delta b_{R}^{PS}+
\frac{3}{10}\left(\frac{2}{5}\Delta b_{B-L}^{LR}
+\frac{3}{5}\Delta b_{R}^{LR}\right)&<17
\end{align}
However, as fig. ~(\ref{fig:MaxDbPS}) shows, Max($\Delta b^{PS}_4$) is a 
rather strong function of the choice of $\Delta b^{LR}_3$. Note, that 
if $m_{PS}$ is low, say below $10^{10}$ GeV larger $\Delta b^{LR}_3$ 
are possible, up to $\Delta b^{LR}_3=7$, see fig. (\ref{fig:MaxDbPS}).
The large values of 
Max($\Delta b^{LR}$) and  Max($\Delta b^{PS}$) allow, in principle, 
a huge number of variants to be constructed in class-II. This is 
demonstrated in fig.  ~(\ref{fig:ScanConfigsPS}), where we show the number 
of variants for an assumed $m_R \sim 1$ TeV as a function of the 
scale $m_{PS}$. Up to $m_{PS} = 10^{15}$ GeV the 
list is exhaustive. For larger values of $m_{PS}$ we have only 
scanned a finite (though large) set of possible variants. 
Note, that these are variants, not configurations. As in the 
case of class-I practically any variant can be made by several 
possible anomaly-free configurations. The exhaustive list of 
variants ($m_{PS} = 10^{15}$ GeV) contains a total of 105909 
possibilities and can be found in \cite{wp}. 

With such a huge number of possible variants, we can discuss only some 
general features here. First of all, within the exhaustive set up to 
$m_{PS}=10^{15}$ GeV, there are a total of 1570 different sets of 
$\Delta b^{LR}_i$, each of which can be completed by more than 
one set of $\Delta b^{PS}_i$. Variants with the same set of 
$\Delta b^{LR}_i$ but different completion of $\Delta b^{PS}_i$ 
have, of course, the same configuration in the LR-regime, but come 
with a different value for $m_{PS}$ for fixed $m_{R}$. Thus, 
they have in general different values for $\alpha_{B-L}$ and 
$\alpha_{R}$ at the LR scale and, see next section, different 
values of the invariants. For example, for the smallest values of 
$\Delta b^{LR}_i$, that are possible in principle [$\Delta b^{LR}_i=
(0,0,1,3/2)$], there are 342 different completing sets of $\Delta b^{PS}_i$.

The very simplest set of $\Delta b^{LR}_i$ possible, $\Delta
b^{LR}_i=(0,0,1,3/2)$, corresponds to the configuration
$\Phi_{1,1,2,-1}+\bar{\Phi}_{1,1,2,-1}$. These fields are necessary to
break $SU(2)_R \times U(1)_{B-L}\to U(1)_Y$. Their presence in the LR
regime requires that in the PS-regime we have at least one set of
copies of $\Psi_{4,1,2}+\bar{\Psi}_{4,1,2}$. In addition, for
breaking the PS group to the LR group, we need at least one copy of
$\Psi_{15,1,1}$. However, the set of $\Psi_{4,1,2} +\bar{\Psi}_{4,1,2}
+\Psi_{15,1,1}$ is not sufficient to generate a sliding
scale mechanism and the simplest configuration that can do so,
consistent with $\Delta b^{LR}_i=(0,0,1,3/2)$, is $3\Psi_{1,2,2}+4
\Psi_{1,1,3}+\Psi_{4,1,2}+\bar{\Psi}_{4,1,2} +\Psi_{15,1,1}$,
leading to $\Delta b^{PS}_i=(6,3,15)$ and a very low possible value of
$m_{PS}$ of $m_{PS}=8.2$ TeV for $m_{R}=1$ TeV (see, however, the
discussion on leptoquarks below). The next possible completion for
$\Phi_{1,1,2,-1}+\bar{\Phi}_{1,1,2,-1}$ is $3\Psi_{1,2,2}+5
\Psi_{1,1,3} +\Psi_{4,1,2}+\bar{\Psi}_{4,1,2}+\Psi_{15,1,1}$,
with $\Delta b^{PS}_i=(6,3,17)$ and $m_{PS}=1.3 \times 10^{8}$ GeV
(for $m_{R}=1$ TeV), etc.

As noted already in section \ref{subsect:LRm}, one copy of 
$\Phi_{1,2,2,0}$ is not sufficient to produce a realistic CKM 
matrix at tree-level. Thus, the minimal configuration of 
$\Phi_{1,1,2,-1}+\bar{\Phi}_{1,1,2,-1}$ relies on the possibility 
of generating all of the departure of the CKM matrix from unity 
by flavor violating soft masses \cite{Babu:1998tm}. There are 
at least two possibilities to generate a non-trivial CKM at tree-level, 
either by adding (a) another $\Phi_{1,2,2,0}$ plus (at least) one copy 
of $\Phi_{1,1,3,0}$ or via (b) one copy of ``vector-like quarks'' 
$\Phi_{3,1,1,\frac{4}{3}}$ or $\Phi_{3,1,1,-\frac{2}{3}}$. 
Consider the configuration $\Phi_{1,1,2,-1}+\bar{\Phi}_{1,1,2,-1}+ 
\Phi_{1,2,2,0}+\Phi_{1,1,3,0}$ first. It leads to 
$\Delta b^{LR}_i=(0,1,4,3/2)$. Since $\Phi_{1,2,2,0}$ and $\Phi_{1,1,3,0}$ 
must come from $\Psi_{1,2,2}$ (or $\Psi_{15,2,2}$) and $\Psi_{1,1,3}$, 
respectively, the simplest completion for this set of $\Delta b^{LR}_i$ 
is again $3\Psi_{1,2,2}+4 \Psi_{1,1,3}+\Psi_{4,1,2}+\bar{\Psi}_{4,1,2}
+\Psi_{15,1,1}$, leading to $\Delta b^{PS}_i=(6,3,15)$ and value of 
$m_{PS}$ of, in this case, $m_{PS}=5.4$ TeV for $m_{R}=1$ TeV. Again, 
many completions with different $\Delta b^{PS}_i$ exist for this 
set of $\Delta b^{LR}_i$.

The other possibility for generating CKM at tree-level, adding for
example a pair of $\Phi_{3,1,1,-\frac{2}{3}} +
\bar{\Phi}_{3,1,1,-\frac{2}{3}}$, has $\Delta b^{LR}_i=(1,0,1,5/2)$
and its simplest PS-completion is $4\Psi_{1,2,2}+4
\Psi_{1,1,3}+\Psi_{4,1,2}+\bar{\Psi}_{4,1,2}
+\Psi_{6,1,1}+\Psi_{15,1,1}$, with $\Delta b^{PS}_i=(7,4,16)$ and a
$m_{PS}=4.6\times 10^6$ TeV for $m_{R}=1$ TeV. Also in this 
case one can find sets with very low values of $m_{PS}$. For example, 
adding a $\Phi_{1,2,2,0}$ to this LR-configuration (for a 
$\Delta b^{LR}_i=(1,1,2,5/2)$), one finds that with the same 
 $\Delta b^{PS}_i$ now a value of $m_{PS}$ as low as $m_{PS}=8.3$ TeV 
for $m_{R}=1$ TeV is possible. 

We note in passing that the original PS-class model of
\cite{DeRomeri:2011ie} in our notation corresponds to 
$\Delta b^{LR}_i=(1,2,10,4)$ and $\Phi_{1,1,2,-1}+\bar{\Phi}_{1,1,2,-1}+ 
\Phi_{1,2,1,1}+\bar{\Phi}_{1,2,1,1}+ 
\Phi_{1,2,2,0}+4 \Phi_{1,1,3,0}+\Phi_{3,1,1,-\frac{2}{3}} +
\bar{\Phi}_{3,1,1,-\frac{2}{3}}$, completed by 
$\Delta b^{PS}_i=(9,5,13)$ with 
$\Psi_{4,1,2}+\bar{\Psi}_{4,1,2}+
\Psi_{4,2,1}+\Psi_{4,2,1}+\Psi_{1,2,2}+4\Psi_{1,1,3}+
\Psi_{6,1,1}+\Psi_{15,1,1}$. The lowest possible $m_{PS}$ 
for a $m_{R}=1$ TeV is $m_{PS}=2.4 \times 10^8$ GeV. Obviously this 
example is not the simplest construction in class-II. We also mention 
that while for the $\beta$-coefficients it does not make any difference, 
the superfield $\Phi_{1,1,3,0}$ can be either interpreted as ``Higgs'' 
or as ``matter''. In the original construction \cite{DeRomeri:2011ie} 
this ``arbitrariness'' was used to assign the 4 copies of 
$\Phi_{1,1,3,0}$ to one copy of $\Omega^c=\Phi_{1,1,3,0}$, i.e. 
``Higgs'' and three copies of $\Sigma^c=\Phi_{1,1,3,0}$, i.e. 
``matter''. In this way $\Omega^c$ can be used to generate the 
CKM matrix at tree-level (together with the extra bi-doublet $\Phi_{1,2,2,0}$), 
while the $\Sigma^c$ can be used to generate an inverse seesaw type-III 
for neutrino masses.

As fig. (\ref{fig:ScanConfigsPS}) shows, there are more than 600 
variant in which $m_{PS}$ can, in principle, be lower than $m_{PS} 
=10^3$ TeV. Such low PS scales, however, are already constrained 
by searches for rare decays, such as $B_s \to \mu^+\mu^-$. This is 
because the $\Psi_{15,1,1}$, which must be present in all our constructions 
for the breaking of the PS group, contains two leptoquark states. 
We will not study in detail leptoquark phenomenology \cite{Davidson:1993qk} 
here, but mention that in the recent paper \cite{Kuznetsov:2012ad} 
absolute lower bounds on leptoquarks within PS models of the order 
of $m_{PS} \simeq 40$ TeV have been derived. There are 426 variants 
for which we find $m_{PS}$ lower than this bound, if we put $m_{R}$ 
to 1 TeV. Due to the sliding scale nature of our construction this, of 
course, does not mean that these models are ruled by the lower 
limit found in \cite{Kuznetsov:2012ad}. Instead, for these models 
one can calculate a lower limit on $m_{R}$ from the requirement 
that $m_{PS}= 40$ TeV. Depending on the model, lower limits on 
$m_{R}$ between $m_{R} = [1.3,27.7]$ TeV are found for the 426 
variants from this requirement.

\begin{figure}[htb]
\centering
\begin{tabular}{cc}
\includegraphics[width=0.45\linewidth]{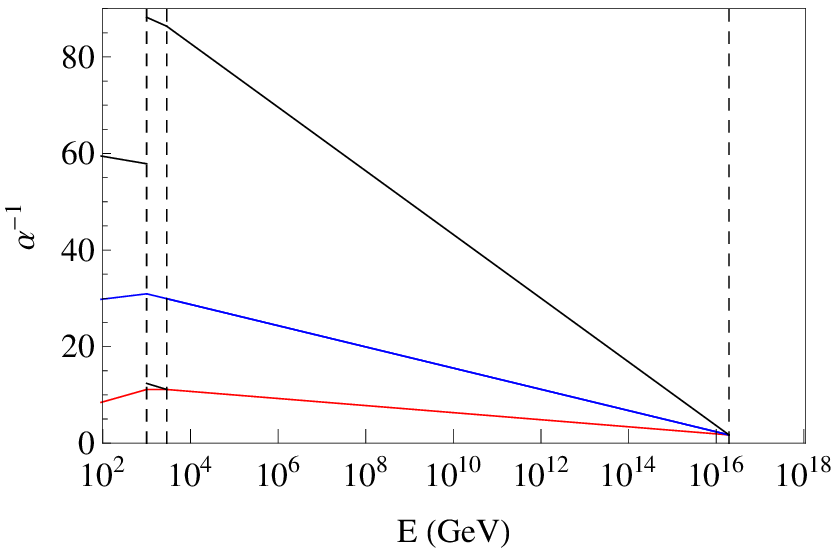}&
\includegraphics[width=0.45\linewidth]{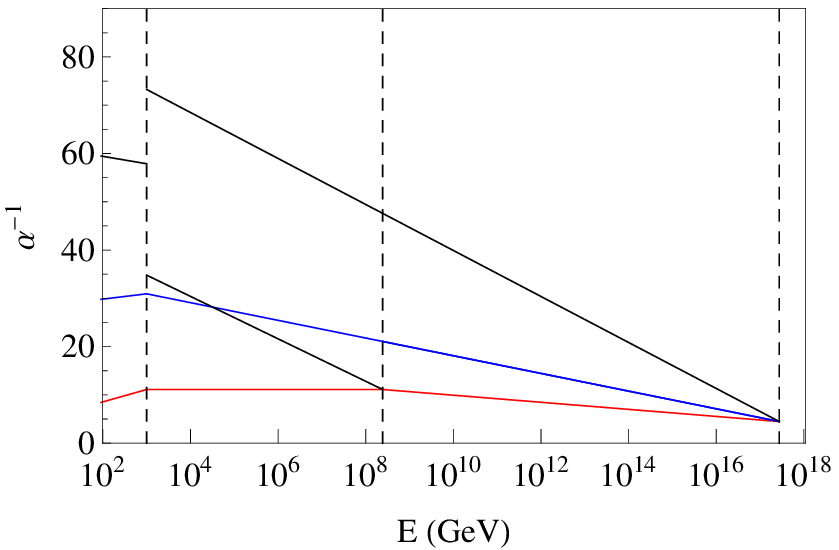}
\end{tabular}
\caption{Gauge coupling unification for PS models with $m_{R}=10^{3}$ GeV. 
In the plot to the left $(\Delta b^{LR}_{3}, \Delta b^{LR}_{L}, \Delta b^{LR}_{R}, 
\Delta b^{LR}_{B-L}, \Delta b^{PS}_{4}, \Delta b_{L}^{PS}, \Delta b_{R}^{PS})
=(3,5,10,3/2,8,5,17$), while the plot to the right corresponds to 
$\Delta b^{'}s=(3,4,12,6,8,4,12)$.}
\label{fig:LowPS_Scale}
\end{figure}

Two example solutions can be seen in fig.~\ref{fig:LowPS_Scale}. 
We have chosen one example with a very low $m_{PS}$ (left) and 
one with an intermediate $m_{PS}$ (right). Note, that different 
from the class-I models, in the class-II models the GUT scale 
is no longer fixed to the MSSM value $m_G \approx 2 \times 10^{16}$ 
GeV. Our samples are restricted to variants which have 
$m_G$ in the interval [$10^{16},10^{18}$] GeV. 

\subsection{Models with an $U(1)_R\times U(1)_{B-L}$ intermediate scale}

Finally, we consider models where there is an additional intermediate
symmetry $U(1)_{R} \times U(1)_{B-L}$ that follows the stage
$SU(2)_{R} \times U(1)_{B-L}$. The field content relevant to this
model is specified in table~\ref{tab:List_of_LR_fields_U1} of the
appendix. In this case the original $SO(10)$ is broken down to the
MSSM in three steps,
\begin{eqnarray}
SO(10) & \to & SU(3)_{c} \times SU(2)_{L} \times SU(2)_{R} \times U(1)_{B-L} 
\\ \nonumber
      & \to &  SU(3)_{c} \times SU(2)_{L} \times U(1)_{R} \times U(1)_{B-L}
        \to  \hskip2mm {\rm MSSM}\textrm{.}
\end{eqnarray}
The first step is achieved in the same way as in class-I models.  The
subsequent breaking $SU(2)_{R}\times U(1)_{B-L}\rightarrow
U(1)_{R}\times U(1)_{B-L}$ is triggered by $\Phi_{5}=\Phi_{1,1,3,0}$
and the last one requires $\Phi_{4}^{'}=\Phi'_{1,1,\frac{1}{2},-1}$,
$\Phi_{20}^{'}=\Phi'_{1,1,1,-2}$ or their conjugates.\\

\noindent 
In theories with more that one $U(1)$ gauge factor, the one loop
evolution of the gauge couplings and soft-SUSY-breaking terms are affected
by the extra kinetic mixing terms. The couplings are defined by the
matrix
\begin{align} G = \left( \begin{array}{cc}
g_{RR} & g_{RX}\\
g_{XR} & g_{XX} \end{array} \right).
\end{align}
and $A(t)=(GG^{T})/(4\pi)=(A^{-1}(t_{0})-\gamma(t-t_{0}))^{-1}$, where
$t = \frac{1}{2\pi}\log(\frac{\mu}{\mu_{0}})$
\cite{DeRomeri:2011ie}. Here, $\mu$ and $\mu_0$ stand for the energy
scale and its normalization point and $A$ is the generalization of
$\alpha$ to matrix form. The matrix of anomalous dimension, $\gamma$,
is defined by the charges of each chiral superfield $f$ under
$U(1)_{R}$ and $U(1)_{B-L}$:
\begin{align}
\gamma = \sum_{f}Q_{f}Q_{f}^{T} 
\end{align}
where $Q_{f}$ denotes a column vector of those charges. Taking the
MSSM's field content we find 
\begin{align}
\gamma= \left( \begin{array}{cc}
7  & 0\\
0  & 6 \end{array} \right).
\end{align}
To ensure the canonical normalization of the $B-L$ charge
within the $SO(10)$ framework, $\gamma$ should be normalized as
$\gamma ^{can}=N\gamma ^{phys}N$, where $N=\text{diag}(1,\sqrt{3/8})$.

\vspace{3mm}

Then, the additional $\beta$ coefficients for the running step $[m_{B-L},
m_{R}]$ are given by,
\begin{align}
(b_{3}^{B-L},b_{2}^{B-L},\gamma_{RR}^{B-L},\gamma_{XR}^{B-L}, \gamma_{XX}^{B-L} ) 
&= (-3,1,6,0, 7 )+
(\Delta b_{3}^{B-L},\Delta b^{B-L}_{2},\Delta \gamma_{RR} ,\Delta \gamma_{XR}, 
\Delta \gamma_{XX}) .
\end{align}

As in the previous PS case, we consider $m_{B-L}=10^{3}$ GeV, $m_{G}
\ge 10^{16}$ GeV and $m_{R}\leq 10^{6}$ GeV. Taking into account the
matching condition:
\begin{align}
 p_{Y}^{T}\cdot A^{-1}(m_{B-L})\cdot p_{Y} = \alpha^{-1}_{1}(m_{B-L})
\end{align}
and $p_{Y}^{T}=(\sqrt{\frac{3}{5}},\sqrt{\frac{2}{5}})$, 
the bounds on the $\Delta b$ are,
\begin{align}
\Delta b^{LR}_{2} & + \frac{3}{10}\Delta b^{B-L}_{2} < 7.1 ,\\
\Delta b^{LR}_{3} & + \frac{3}{10}\Delta b^{B-L}_{3} < 6.9 , \nonumber\\
\frac{3}{5} \Delta b^{LR}_{R} + \frac{2}{5} \Delta b^{LR}_{B-L} + 
&  \frac{3}{10}p_{Y}^{T} \cdot \Delta \gamma \cdot p_{Y} < 10.8 .\nonumber
\end{align}

\begin{figure}[htb]
\centering
\begin{tabular}{cc}
\includegraphics[width=0.45\linewidth]{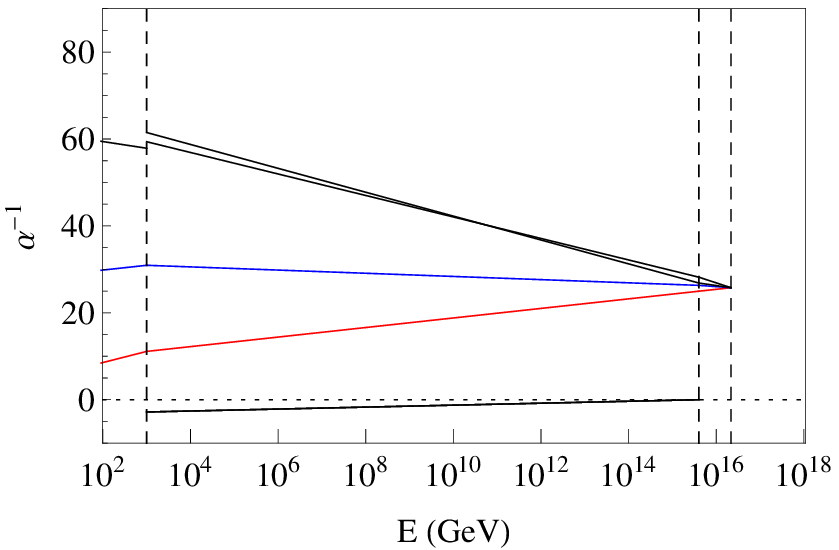}&
\includegraphics[width=0.45\linewidth]{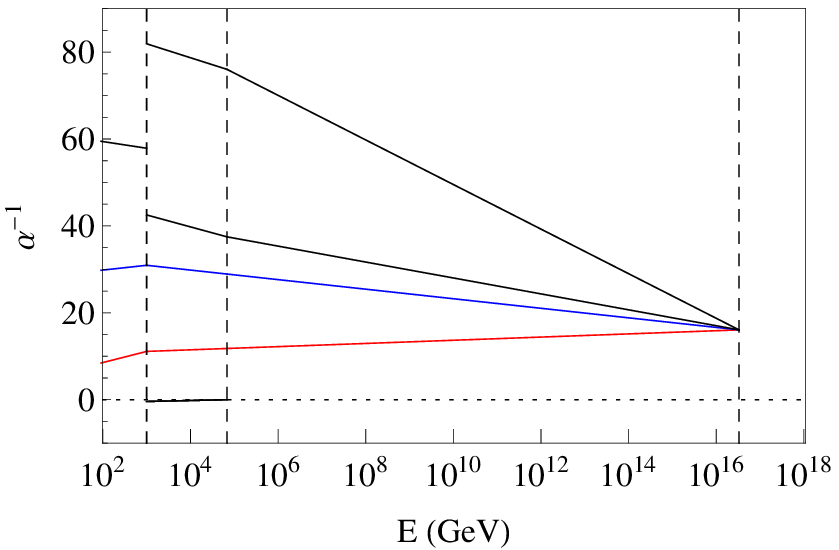}
\end{tabular}
\caption{Gauge coupling unification in models with an $U(1)_R\times
  U(1)_{B-L}$ intermediate scale, for $m_{R}=10^{3}$ GeV.  Left: $(\Delta
  b^{LR}_{3}, \Delta b^{LR}_{L}, \Delta b^{LR}_{R}, \Delta
  b^{LR}_{B-L}, \Delta b^{B-L}_{3}, \Delta b_{L}^{B-L}, \Delta
  \gamma_{RR}, \Delta \gamma_{XR}, \Delta
  \gamma_{XX})=(0,1,3,3,0,0, 1/2, -\sqrt{3/8}, 3/4)$. Right: 
($2,2,4,8,2,2,1/2,-\sqrt{3/8},11/4$). The line, which appears 
close to zero in the $U(1)_R\times U(1)_{B-L}$ regime is the running 
of the off-diagonal element of the matrix $A^{-1}$, i.e. measures the 
size of the $U(1)$-mixing in the model.}
\label{fig:U1MixingLowScale}
\end{figure}

Even with this restriction in the scales we found 15610 solutions,
more than in the PS case, due to the fact that there are more $\Delta
b^{'}s$ that can be varied to obtain solutions.  The qualitative
features of the running of the gauge couplings are shown for two 
examples in fig.~(\ref{fig:U1MixingLowScale}). In those two examples 
the $(\Delta b^{LR}_{3}, \Delta b^{LR}_{L}, \Delta b^{LR}_{R}, 
\Delta b^{LR}_{B-L}, \Delta b^{B-L}_{3}, \Delta b_{L}^{B-L}, \Delta \gamma_{RR}, 
\Delta \gamma_{XR}, \Delta \gamma_{XX})$ have been chosen as 
$(0,1,3,3,0,0, 1/2, -\sqrt{3/8}, 3/4)$ (left) and 
($2,2,4,8,2,2,1/2,-\sqrt{3/8},11/4$) (right). The former corresponds 
to the minimal configuration $\Phi'_{1,1,1/2,-1}+\bar{\Phi}'_{1,1,1/2,-1}$ 
in the lower regime and $\Phi_{1,1,2,-1}+\bar{\Phi}_{1,1,2,-1}+\Phi_{1,1,3,0}+
\Phi_{1,2,1,1}+\bar{\Phi}_{1,2,1,1}$ in the higher (LR-symmetric regime). 
The latter corresponds to 
$\Phi'_{1,1,1/2,-1}+\bar{\Phi}'_{1,1,1/2,-1}+\Phi'_{1,3,0,0}+2 \Phi'_{3,1,1,-2/3}+
2 \bar{\Phi}'_{3,1,1,-2/3}$  and 
$2(\Phi_{1,1,2,-1}+\bar{\Phi}_{1,1,2,-1})+\Phi_{1,1,3,0}++\Phi_{1,3,1,0}
+\Phi_{1,1,1,2}+\bar{\Phi}_{1,1,1,2}+2(\Phi_{3,1,1,-2/3}+ \bar{\Phi}_{3,1,1,-2/3})$, 
respectively.

For models in this class, the sliding condition requires that the
unification scale is independent of $m_{B-L}$ and this happens when
\begin{alignat}{1}
0= & \begin{pmatrix}\Delta b_{3}^{B-L}-\Delta b_{2}^{B-L} , & p_{Y}^{T}\cdot\Delta\gamma\cdot p_{Y}-\Delta b_{2}^{B-L}\end{pmatrix}.\begin{pmatrix}\begin{array}{rr}
0 & 1\\
-1 & 0
\end{array}\end{pmatrix}.\begin{pmatrix}\Delta b_{3}^{LR}-\Delta b_{2}^{LR}\\
\frac{3}{5}\Delta b_{R}^{LR}+\frac{2}{5}\Delta b_{B-L}^{LR}-\Delta b_{2}^{LR}-\frac{18}{5}
\end{pmatrix}.
\end{alignat}
Similarly to PS models, in this class of models the higher
intermediate scale ($m_{R}$) depends, in general, on the lower one
($m_{B-L}$). However, there is also here a special condition which
makes both $m_{R}$ and $m_{G}$ simultaneously independent of $m_{B-L}$,
which is
\begin{align}
\Delta b_{3}^{LR} = \Delta b_{2}^{LR} =  p_{Y}^{T} \cdot \Delta \gamma \cdot p_{Y}.
\end{align}
Models of this kind are, for example, those with $\Delta b_{3} = 0$
and $m_{R} $ large, namely $m_{R} \geq 10^{13}$ GeV. One case is given
by the model in \cite{DeRomeri:2011ie}, where $m_{R} \simeq 4 \times
10^{15}$ GeV.  

\section{Invariants}
\label{sect:invariants}

\subsection{Leading-Log RGE Invariants}
\label{sect:leadloginv}

In this section we briefly recall the basic definitions
\cite{DeRomeri:2011ie} for the calculation of the ``invariants''
\cite{Buckley:2006nv,Hirsch:2008gh,Esteves:2010ff}.  In mSugra there
are four continuous and one discrete parameter: The common gaugino
mass $M_{1/2}$, the common scalar mass $m_{0}$, the trilinear coupling
$A_{0}$ and the choice of the sign of the $\mu$-parameter,
sgn($\mu$). In addition, the ratio of vacuum expectation values of
$H_{d}$ and $H_{u}$, $\tan \beta=\frac{v_{u}}{v_{d}}$ is a free
parameter. The latter is the only one defined at the weak scale, while
all the others are assigned a value at the GUT scale.

Gaugino masses scale as gauge couplings do and so the requirement of
GCU fixes the gaugino masses at the low scale
\begin{eqnarray}
M_i(m_{SUSY}) = \frac{\alpha_i(m_{SUSY})}{\alpha_G} M_{1/2}.
\label{eq:gaugino}
\end{eqnarray}
Neglecting the Yukawa and soft trilinear couplings for the soft mass 
parameters of the first two generations of sfermions one can write
\begin{alignat}{1}\label{eq:scalar}
m_{\tilde{f}}^{2}-m_{0}^{2} & = \frac{M_{1/2}^{2}}{2\pi\alpha_G^{2}}
  \sum_{R_{j}}\sum_{i=1}^{N}c_{i}^{f,R_{j}}
  \alpha_{i-}^{R_{j}}\alpha_{i+}^{R_{j}}
  \left(\alpha_{i-}^{R_{j}}+\alpha_{i+}^{R_{j}}\right)
   \log\frac{m_{+}^{R_{j}}}{m_{-}^{R_{j}}}\,.
\end{alignat}
Here, the sum over ``$R_j$'' runs over the different regimes in the
models under consideration, while the sum over $i$ runs over all gauge
groups in a given regime.  $m_{+}^{R_{j}}$ and $m_{-}^{R_{j}}$ are the
upper and lower boundaries of the $R_{j}$ regime and
$\alpha_{i+}^{R_{j}}$, $\alpha_{i-}^{R_{j}}$ are the values of the
gauge coupling of group $i$, $\alpha_{i}$, at these scales.  As for
the coefficients $c_i$, they can be calculated from the quadratic
Casimir of representations of each field under each gauge group $i$
and are given for example in \cite{DeRomeri:2011ie}. In the presence
of multiple U(1) gauge groups the RGEs are different (see for
instance \cite{Fonseca:2011vn} and references contained therein) and
this leads to a generalization of equation (\ref{eq:scalar}) for the
U(1) mixing phase \cite{DeRomeri:2011ie}.  Here we just quote the end
result (with a minor correction to the one shown in this last
reference) ignoring the non-U(1) groups:
\begin{alignat}{1}\label{eq:m2f_U1mixing}
\tilde{m}_{f-}^{2}-\tilde{m}_{f+}^{2} & =
\frac{M_{1/2}^{2}}{\pi\alpha_{G}^{2}}
    Q_{f}^{T}A_{-}\left(A_{-}+A_{+}\right)A_{+}Q_{f}
   \log\frac{m_{+}}{m_{-}},
\end{alignat}
where $m_{+}$ and $m_{-}$ are the boundary scales of the U(1) mixing regime 
and $A_{+}$, $A_{-}$ are the $A$ matrix defined in the previous section 
(which generalizes $\alpha$) evaluated in these two limits. Likewise,
$\tilde{m}_{f+}^{2}$ and $\tilde{m}_{f-}^{2}$ are the values of the
soft mass parameter of the sfermion $\tilde{f}$ at these two energy
scales. The equation above is a good approximation to the result obtained by 
integration of the following 1-loop RGE for the soft masses which assumes 
unification of gaugino masses and gauge coupling constants:
\begin{alignat}{1}
\frac{d}{dt}\tilde{m}_{f}^{2} & =-\frac{4M_{1/2}^{2}}{\alpha_{G}^{2}}Q_{f}^{T}A^{3}Q_{f}.\label{eq:DerivariveOfm2f}
\end{alignat}
Note that in the limit where the U(1) mixing phase extends all the way
up to $m_{G}$, the $A$ matrix measured at different energy scales will
always commute and therefore equation (\ref{eq:m2f_U1mixing})
presented here matches the one in \cite{DeRomeri:2011ie} and in
fact both are exact integrations of
(\ref{eq:DerivariveOfm2f}). However, if this is not the case, it is
expected that there will be a small discrepancy between the two
approximations, which nevertheless is numerically small and therefore
negligible.

From the five soft sfermion mass parameters of the MSSM and one of the
gaugino masses it is possible to form four different combinations
that, at 1-loop level in the leading-log approximation, do not depend
on the values of $m_{0}$ and $M_{1/2}$ and are therefore called invariants:
\begin{eqnarray}
\label{eq:definv}
LE &= ({m_{\widetilde{L}}^{2}-m_{\widetilde{E}}^{2}})/{M_{1}^{2}},  \\
QE &= ({m_{\widetilde{Q}}^{2}-m_{\widetilde{E}}^{2}})/{M_{1}^{2}}, \nonumber 
\\
DL &= ({m_{\widetilde{D}}^{2}-m_{\widetilde{L}}^{2}})/{M_{1}^{2}}, \nonumber 
\\
QU &= ({m_{\widetilde{Q}}^{2}-m_{\widetilde{U}}^{2}})/{M_{1}^{2}}. \nonumber
\end{eqnarray}
While being pure numbers in the MSSM, invariants depend on the 
particle content and gauge group in the intermediate stages, 
as shown by eq. (\ref{eq:scalar}). 

We will not discuss errors in the calculation of the invariants 
in detail, we refer the interested reader to \cite{DeRomeri:2011ie} 
and for classical $SU(5)$ based SUSY seesaw models to 
\cite{Hirsch:2008gh,Esteves:2010ff}.

We close this subsection by discussing that not all model variants
which we presented in section \ref{sec:models} will be testable by
measurements involving invariants at the LHC. According to
\cite{Baer:2012vr} the LHC at $\sqrt{s}=14$ TeV will be able to
explore SUSY masses up to $m_{\tilde g}\sim 3.2$ TeV ($3.6$ TeV) for
$m_{\tilde q} \simeq m_{\tilde g}$ and of $m_{\tilde g}\sim 1.8$ TeV
($2.3$ TeV) for $m_{\tilde q} \gg m_{\tilde g}$ with 300 fb$^{-1}$
(3000 fb$^{-1}$). The LEP limit on the chargino, $m_\chi> 105$
GeV~\cite{Beringer:1900zz}, translates into a lower bound for
$M_{1/2}$, with the value depending on the $\Delta b$. For the class-I
models with $\Delta b = 5$ this leads to $M_{1/2} \gsim 1.06$
TeV. One can assume conservatively $m_0=0$ GeV and calculate from this
lower bound on $M_{1/2}$ a lower limit on the expected squark masses
in the different variants. All variants with squark masses above the
expected reach of the LHC-14 will then not be testable via
measurements of the invariants. This discards all models with $\Delta
b = 5$ as untestable unfortunately.

For completeness we mention that if we take the present LHC limit on
the gluino, $m_{\tilde g} \gsim 1.1$ TeV~\cite{ATLAS-CONF-2012-109}, this will
translate into a lower limit $M_{1/2}\gsim 4.31$ TeV for $\Delta
b=5$. We have also checked that models with $\Delta b = 4$, can still
have squarks with masses testable at LHC, even for the more recent
LHC bound on the gluino mass.

\subsection{Classification for invariants}
\label{sect:invclasses}

\noindent 
The invariants defined in eq. (\ref{eq:definv}) are pure numbers 
in mSugra and receive corrections which can, in principle, either 
be positive or negative once new superfields (and/or gauge groups) are 
added to the MSSM. If we simply consider whether invariants are 
larger or smaller than their respective values in mSugra, with four 
invariants there are in principle $2^4=16$ possibilities. We 
arbitrarily assign each of them a number as listed in table 
\ref{tab:classes}.

\begin{table}[htb]
\centering
\begin{tabular}{|c|c|c|c|c|c|c|c|c|c|c|c|c|c|c|c|c|}\hline
Set \#&1&2&3&4&5&6&7&8&9&10&11&12&13&14&15&16\\\hline
$\Delta LE$&+&+&+&+&+&+&+&+&$-$&$-$&$-$&$-$&$-$&$-$&$-$&$-$\\ \hline
$\Delta QE$&+&+&+&$-$&+&$-$&$-$&$-$&+&+&+&$-$&+&$-$&$-$&$-$\\ \hline
$\Delta DL$&+&+&$-$&+&$-$&+&$-$&$-$&+&+&$-$&+&$-$&+&$-$&$-$\\ \hline
$\Delta QU$&+&$-$&+&+&$-$&$-$&+&$-$&+&$-$&+&+&$-$&$-$&+&$-$\\ \hline
Class-I? & \checked & \checked& & & & & & & &\checked & & & 
&\checked & &  \\ \hline
Class-II? & \checked & \checked &\checked & & & \checked & \checked 
&\checked&  &\checked & & & &\checked & & \checked \\ \hline
Class-III? & \checked & \checked& & & & & & & &\checked & & & 
&\checked & &  \\ \hline
\end{tabular}
\caption{The 16 different combinations of signs for 4 invariants.  We
  assign a ``$+$'' if the corresponding invariant at $m_{SUSY}$ is
  larger than its value in mSugra and ``$-$'' otherwise. As discussed
  in the text, only 9 of the 16 different sign combinations can be
  realized in the models we consider. Moreover, for class-I only the
  sets 1,2, 10 and 14 can be realized, see discussion. For class-III
  we also have found only sets 1,2, 10 and 14, but here our search was
  not exhaustive.}

    \label{tab:classes}
\end{table}

However, it is easy to demonstrate that not all of the 16 sets can 
be realized in the three classes of models we consider. This can be 
understood as follows. If all sfermions have a common $m_0$ at 
the GUT scale, then one can show that
\begin{align}\label{eq:sumrule}
m_{\widetilde{E}}^{2}-m_{\widetilde{L}}^{2}+m_{\widetilde{D}}^{2}-
2 m_{\widetilde{U}}^{2}+m_{\widetilde{Q}}^{2}=0
\end{align}
holds independent of the energy scale, at which soft masses are 
evaluated. This relation is general, regardless of the combination of
intermediate scales that we may consider and for all gauge groups 
we consider. It is a straightforward 
consequence of the charge assignments of the standard model
fermions and can be easily checked by calculating the Dynkin
coefficients of the E,L,D,U and Q representation in the different
regimes. In terms of the invariants, this relation becomes:
\begin{align}\label{eq:sumruleinv}
QE = DL+2QU,
\end{align}
i.e. only three of the four invariants are independent. From eq.
(\ref{eq:sumruleinv}) it is clear that if $\Delta DL$ and $\Delta QU$ are
both positive (negative), then $\Delta QE$ must be also positive
(negative). This immediately excludes the sets 4, 5, 12 and 13.

Within the MSSM group eq. (\ref{eq:sumrule}) allows one relation 
among the invariants. However, one can calculate the relations 
among the Dynkin indices of the MSSM sfermions within the extended 
gauge groups we are considering and in these there is one additional 
relation:
\begin{align}\label{LRsumrule}
QU = LE.
\end{align}
Since eq. (\ref{LRsumrule}) is valid only in the regime(s) with 
extended gauge group(s), it is not exact, once the running 
within the MSSM regime is included. However, taking into account 
the running within the MSSM group one can write: 
\begin{align}\label{LRsumruleApp}
QU = LE+f(m_{R}),
\end{align}
with
\begin{align}
f\left(m_{R}\right)=\frac{2}{33}\left\{ \left[\frac{33}{10\pi}\alpha_{1}^{MSSM}\log\left(\frac{m_{R}}{m_{SUSY}}\right)-1\right]^{-2}-1\right\} 
\end{align}\\
Here, $\alpha_{1}^{\rm MSSM}$ is the value of $\alpha_1$ at $m_{SUSY}$. 
It is easy to see that $f(m_{R})$ is always small ($<0.3$) and 
positive and, vanishes if $m_R$ approaches $m_{SUSY}$.  Note, 
that here $m_R$ stands for the scale where the MSSM group 
is extended, in the class-III models it is therefore $m_{B-L}$. 

Eq. (\ref{LRsumruleApp}) allows to eliminate three more cases from
table \ref{tab:classes}. Since $f(m_{R})$ is positive, $\Delta QU \leq
\Delta LE$ always, so, it is not possible to have $\Delta LE =-$ and
$\Delta QU=+$. This excludes three additional sets from table 
\ref{tab:classes}: 9, 11 and 15, leaving a total of 9 possible sets.

Finally, in class-I models it is possible to eliminate four more 
sets, namely all of those with $\Delta DL<0$. It is easy to see, with the 
help of eq.(\ref{eq:scalar}) that this is the case. It follows 
from the fact that in the LR case, the $c^L_i$ are non-zero for 
$U(1)_{B-L}$ and $SU(2)_L$ with the values $3/4$ and $3/2$, respectively. 
Since also the sum is smaller than the $c^D_3$ (and $\alpha_3$ is 
larger than the other couplings, $D$ must run faster than $L$ in 
the LR-regime.

By the above reasoning set 6 seems to be, in principle, possible in
class-I, but is not realized in our complete scan. We found a few
examples in class-II, see below. Due to the (approximate) relation
QU=LE it seems a particularly fine-tuned situation.  We also note in
passing, that in the high-scale seesaw models of type-II
\cite{Hirsch:2008gh} and seesaw type-III \cite{Esteves:2010ff} with
running only within the MSSM group, all invariants run always towards
larger values, i.e. only set 1 is realized in this case.

The above discussion serves only as a general classification of 
the types of sets of invariants that can be realized in the different 
model classes. The numerical values of the invariants, however, 
depend on both, the variant of the model class and the scale of 
the symmetry breaking. We will discuss one example for each possible set 
next.

\subsection{Invariants in model class-I}

\begin{figure}[h]
\begin{center}
\begin{tabular}{cc}
Set 1 & Set 2 \\
\includegraphics[scale=0.8]{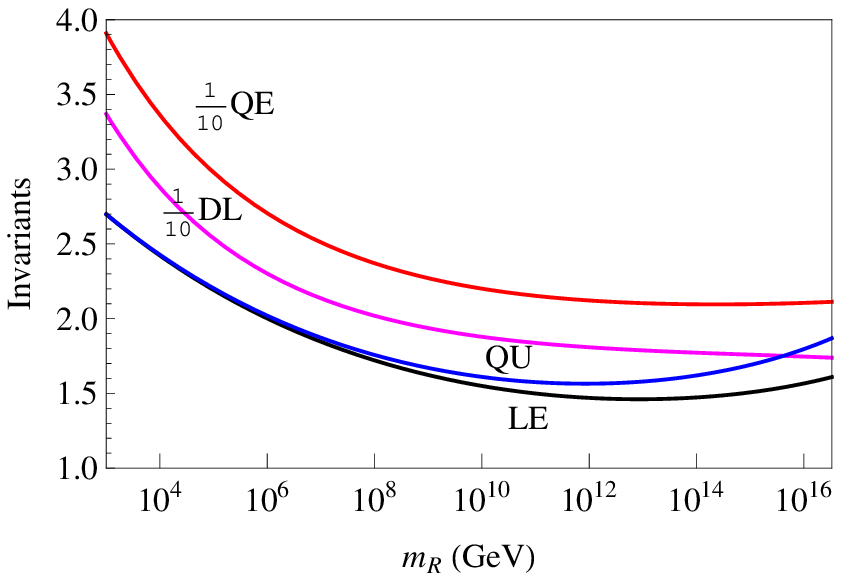} 
&\includegraphics[scale=0.8]{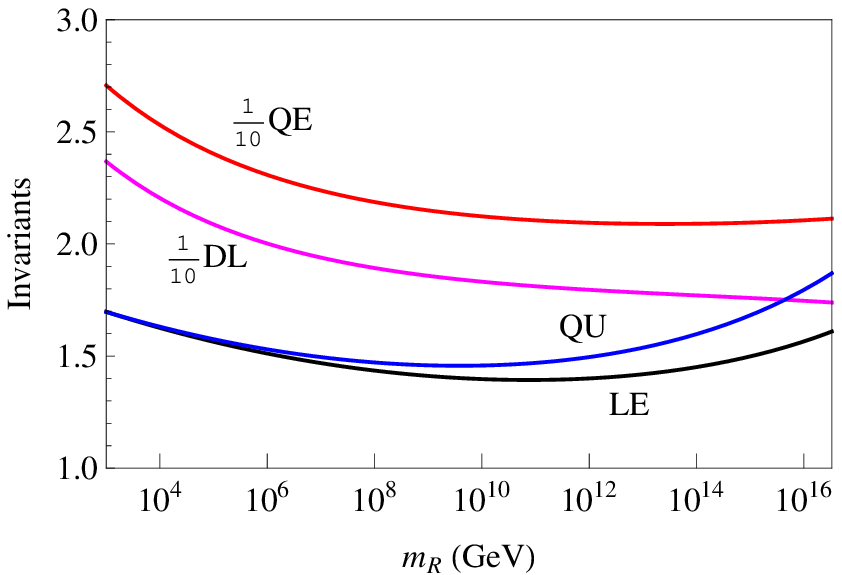}\\
Set 10 & Set 14\\
\includegraphics[scale=0.8]{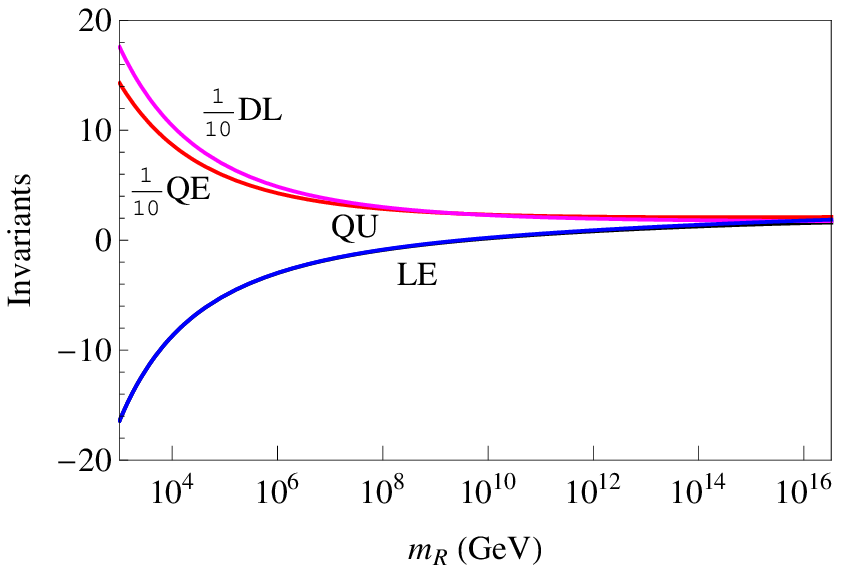} 
&\includegraphics[scale=0.8]{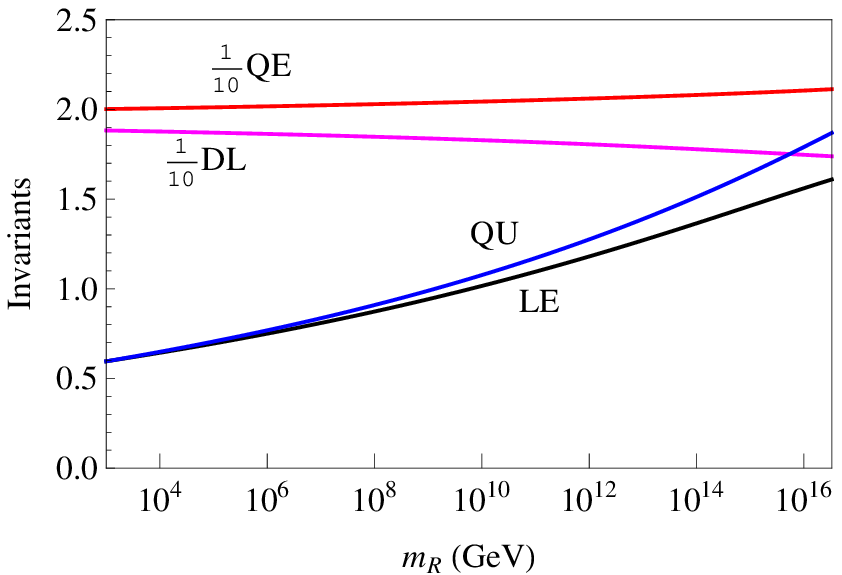}
\end{tabular}
\end{center}
\caption{\label{fig:invLR}$m_{R}$ dependence of the invariants in
model class-I.  The examples of $\Delta b^{LR}_i= (\Delta
b^{LR}_{3},b^{LR}_{L}, \Delta b^{LR}_{R}, \Delta b^{LR}_{BL})$ for
these sets are as follows. Set 1: ($2,2,9,1/2$), Set 2: ($1,1,7,1$),
Set 10: ($4,4,3,29/2$), Set 14: ($0,0,2,6$). For a discussion see text.}
\end{figure}

\noindent 
Fig. (\ref{fig:invLR}) shows examples of the $m_{R}$ dependence 
of the invariants corresponding to the four cases: sets 1, 2, 10 and 
14 of table \ref{tab:classes}. Note that we have scaled down the 
invariants $QE$ and $DL$ for practical reasons. Note also the different 
scales in the different plots.

In all cases QU $\simeq$ LE, if the LR scale extends to very low
energies. As explained above, this is a general feature of the
extended gauge groups we consider and thus, measuring a non-zero QU-LE
allows in our setups, in principle, to derive a lower limit on the
scale at which the extended gauge group is broken.

Sets 1 and 2 show a quite similar overall behavior in these
examples. Set 1, however, can also be found in variants of class-I with
larger $\beta$ coefficients, i.e. larger quantitative changes with
respect to the mSugra values. It is possible to find variants within
class-I which fall into set 2, but again due to the required
similarity of QU and LE, this set can be realized only if both QU and
LE are numerically very close to their mSugra values. 
Set 14 in class-I, finally, is possible only with QE and DL close to 
their mSugra values, as can be understood from eq. (\ref{eq:sumruleinv}). 

In general, for variants with large $\Delta b^{LR}_{3}$ changes in the
invariants can be huge, see for example the plot shown for set 10. The
large change is mainly due to the rapid running of the gaugino masses
in these variants, but also the sfermion spectrum is very ``deformed''
with respect to mSugra expectations. For example, a negative LE means
of course that left sleptons are lighter than right sleptons, a
feature that can never be found in the ``pure'' mSugra model. Recall that
for solutions with $\Delta b^{LR}_3=5$, the value of the squark masses
lies beyond the reach of the LHC.

\subsection{Model class-II}

\begin{figure}[h]
\begin{center}
\begin{tabular}{cc}
Set 3 & Set 6 \\
\includegraphics[scale=0.8]{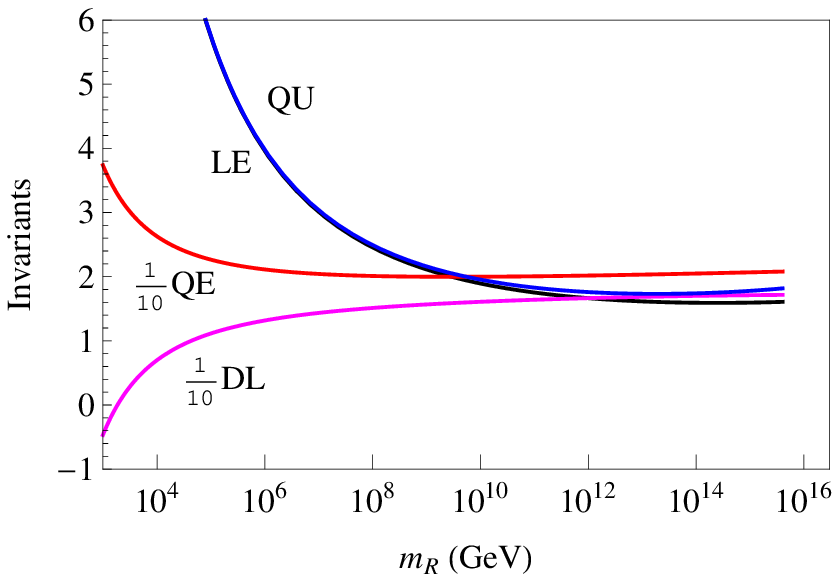} 
&\includegraphics[scale=0.8]{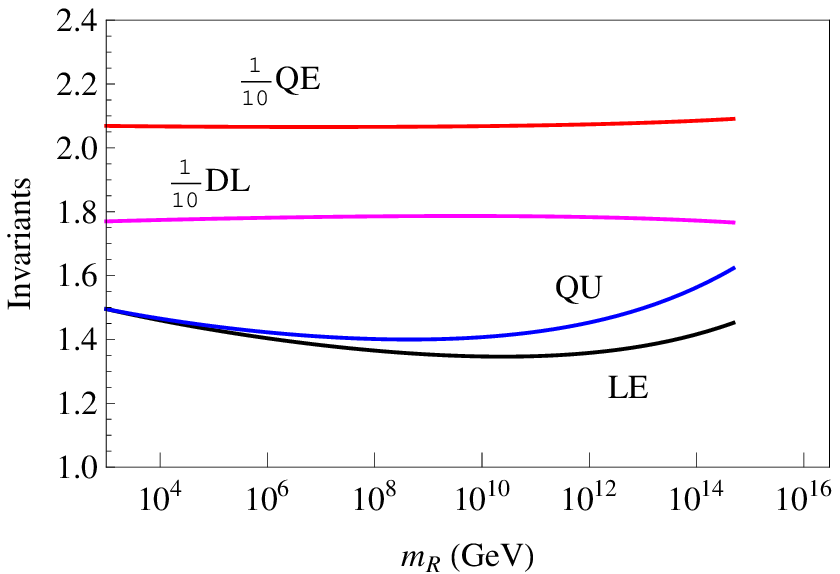}
\end{tabular}
\begin{center}
Set 7 \\
\includegraphics[scale=0.8]{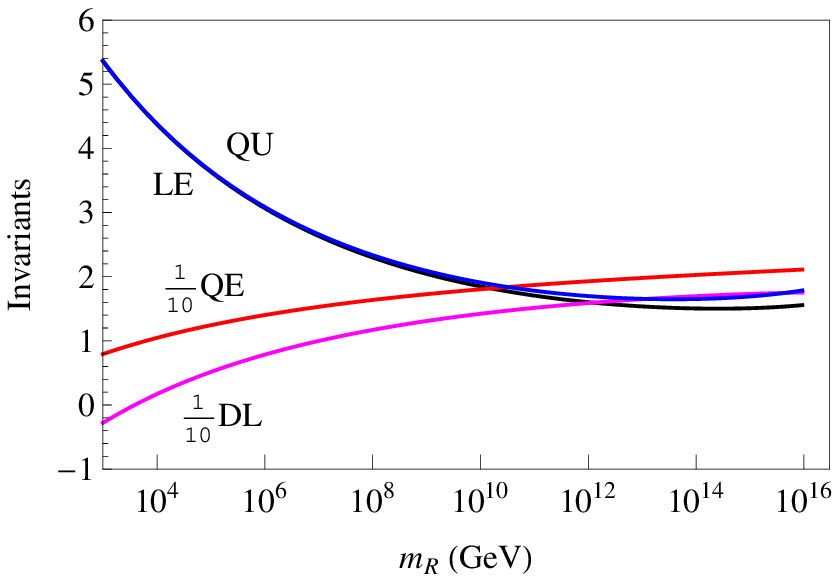}
\end{center}
\begin{tabular}{cc}
Set 8 & Set 16 \\
\includegraphics[scale=0.8]{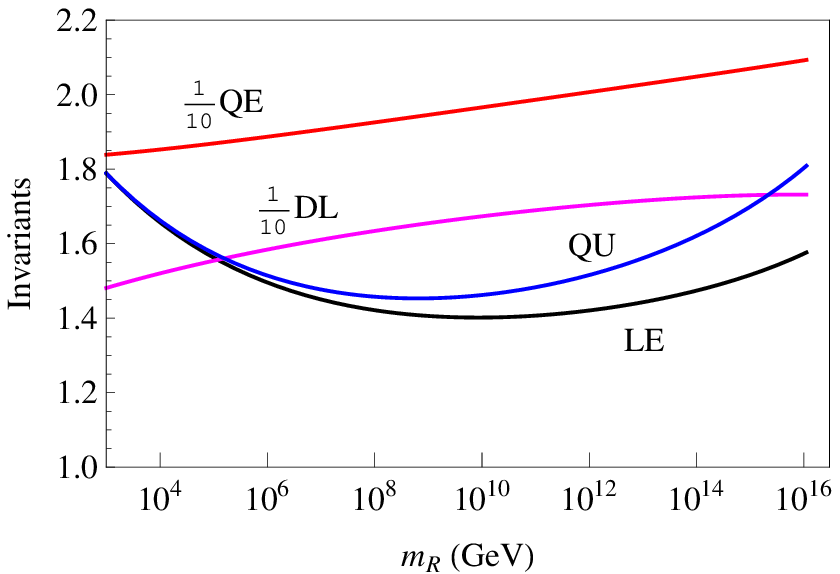} 
& \includegraphics[scale=0.8]{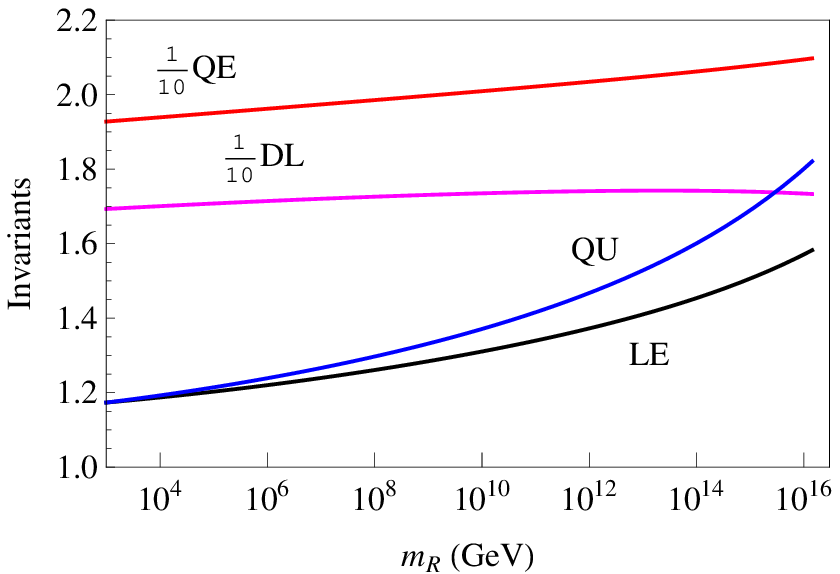} 
\end{tabular}
\end{center}
\caption{\small{\label{fig:invPS}The $m_{R}$ dependence of the
invariants in model class-II. The examples shown correspond to the
choices of $\Delta b=(\Delta b^{LR}_{3},\Delta b^{LR}_{L}, \Delta
b^{LR}_{R}, \Delta b^{LR}_{BL}, \Delta b^{PS}_{4}, \Delta b_{L}^{PS},
\Delta b_{R}^{PS})$: Set 3: ($0,1,10,3/2,14,9,13$), Set 6:
($0,0,1,9/2,63,60,114$), Set 7: ($0,3,12,1.5,6,3,15$), Set 8:
($0,0,9,1.5,11,8,12$), Set 16: ($0,0,7,1.5,11,8,10$). } }
\end{figure}

Fig. (\ref{fig:invPS}) shows examples of the invariants for class-II 
models for those cases of sets, which can not be covered in class-I.
Again, QU and DL are scaled and different plots show differently 
scaled axes.

The example for set 3 shown in fig. (\ref{fig:invPS}) is similar 
to the one of the original prototype model constructed in 
\cite{DeRomeri:2011ie}. For set 6 we have found only a few examples, 
all of them show invariants which hardly change with respect to the 
mSugra values of the invariants. 
The example for set 7 shows that also QE can decrease considerably 
in some variants with respect to its mSugra value. 
Set 8 is quantitatively similar to set 2 and set 16 quite similar 
numerically to set 14. To distinguish these, highly accurate SUSY 
mass measurements would be necessary.

Again we note that larger values of $\Delta b^{LR}$, especially 
large $\Delta b^{LR}_3$, usually lead to numerically larger changes 
in the invariants, making these models in principle easier to test.

\subsection{Model class-III}

\begin{figure}[h]
\begin{center}
\begin{tabular}{cc}
$\Delta b_{3} = 0$ & $\Delta b_{3} \neq 0$ \\
\includegraphics[scale=0.8]{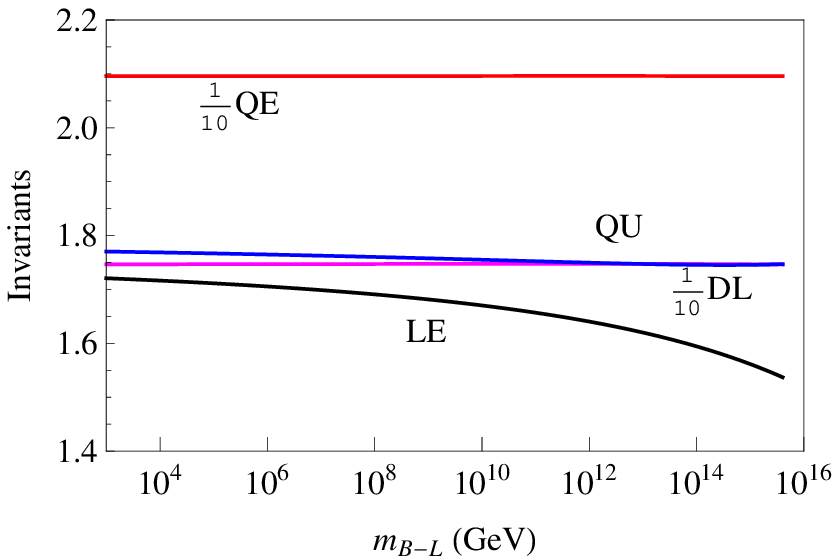} 
& \includegraphics[scale=0.8]{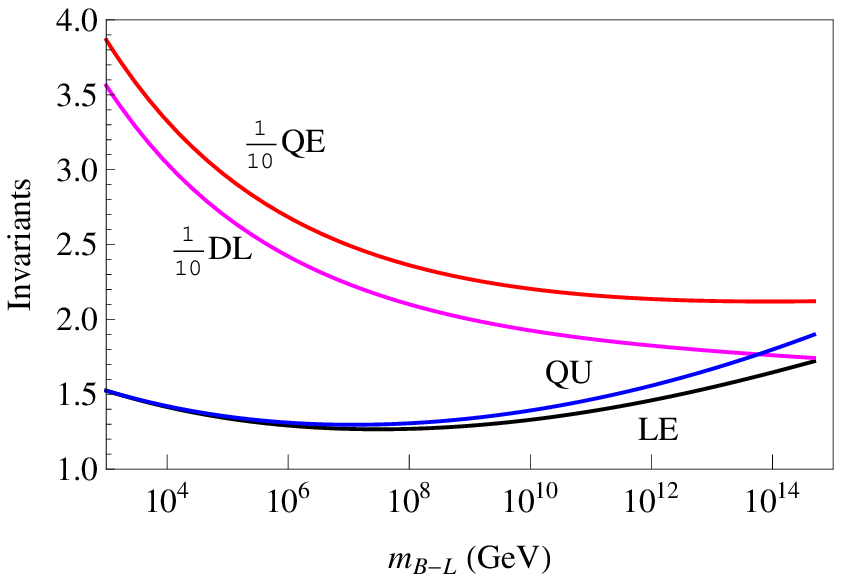}
\end{tabular}
\end{center}
\caption{\small{\label{fig:invBL}The $m_{B-L}$ dependence of the
invariants in Model III. To the left the example chooses: 
$(\Delta b^{LR}_{3}, \Delta
b^{LR}_{L}, \Delta b^{LR}_{R}, \Delta b^{LR}_{BL}, \Delta
b^{BL}_{3}, \Delta b^{BL}_{L}, \Delta \gamma_{RR}, \Delta
\gamma_{XR}, \Delta \gamma_{XX})=(0,1,3,3,0,0, 1/2, -\sqrt{3/8}, 3/4)$. 
To the right: ($2,2,4,8,2,2,1/2,-\sqrt{3/8},11/4$). }}
\end{figure}  

\noindent 
Here, the invariants depend on $m_{B-L}$ with a milder or stronger
dependence, depending on the value of $\Delta b_{3}$. For
almost all the solutions with $\Delta b_{3}=0$ , the values $QU$,
$DL$, $QE$ are constants and only in $LE$ a mild variation with
$m_{B-L}$ is found. This fact was already pointed out in 
\cite{DeRomeri:2011ie}. However, we have found that class-III 
models can be made with $\Delta b_{3}>0$ and these, in general, 
lead to invariants which are qualitatively similar to the case 
of class-I discussed above. In fig. (\ref{fig:invBL}) we show 
two examples of invariants for class-III, one with 
$\Delta b_{3}=0$ and one with $\Delta b_{3}=1$.

The solutions with $\Delta b_{3}\neq 0$ fall in two kinds: The minimum
value of $m_{R}$ is very large. Then, the invariants have the same
behavior than those in which $\Delta b_{3}=0$. The minimum value of
$m_{R}$ is low. The invariants are not constants and look similar to
the ones in the class-I models. The generally mild dependence on $m_{B-L}$ 
can be understood, since it enters into the soft masses only
through the changes in the abelian gauge couplings. Class-III models  
are therefore the hardest to ``test'' using invariants.

\subsection{Comparison of model classes}

\begin{figure}
\begin{centering}
\includegraphics[scale=0.25]{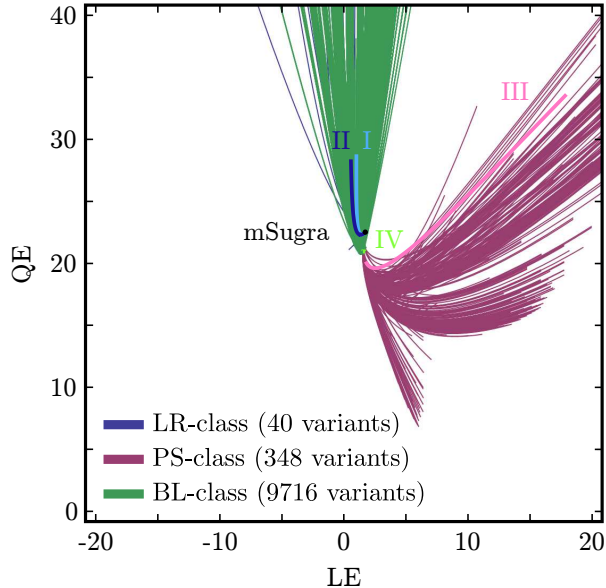}
\par\end{centering}

\caption{\label{fig:LEQE_parametric_plot_RGB}Parametric
$\left(LE,QE\right)$ plot for the different variants (see text). The
thicker lines labeled with I, II, III and IV indicate the result for
the four prototype models presented in \cite{DeRomeri:2011ie}.}
\end{figure}

The classification of variants that we have discussed in section
\ref{sect:invclasses} only takes into account what happens when the
lowest intermediate scale is very low, ${\cal O}(m_{SUSY})$.  When one
varies continuously the lowest intermediate scale ($m_{R}$ in the LR
and PS-class models or $m_{B-L}$ in the BL-class of models), each
variant draws a line in the 4-dimensional space
$\left(LE,QU,DL,QE\right)$. The dimensionality of such a plot can be
lowered if we use the (approximate) relations between the invariants
shown above, namely $QU\approx LE$ and $QE=DL+2QU$. We can then choose
two independent ones, for example $LE$ and $QE$, so that the only
non-trivial information between the 4 invariants is encoded in a
$\left(LE,QE\right)$ plot. In this way, it is possible to simultaneous
display the predictions of different variants. This was done in
fig.~(\ref{fig:LEQE_parametric_plot_RGB}), where LR-, PS- and
BL-variants are drawn together. The plot is exhaustive in the sense
that it includes all LR-variants, as well as all PS- and BL-variants
which can have the highest intermediate scale below $10^{6}$ GeV. In
all cases, we required that $\alpha^{-1}$ at unification is larger
than $\nicefrac{1}{2}$ when the lowest intermediate scale is equal to
$m_{SUSY}$.

There is a dot in the middle of the figure - the mSugra point - which
corresponds to the prediction of mSugra models, in the approximation
used. It is expected that every model will draw a line with one end
close to this point. This end-point corresponds to the limit where
the intermediate scales are close to the GUT scale and therefore the
running in the LR, PS and BL phases is small so the invariants should
be similar to those in mSugra models. So the general picture is that
lines tend to start (when the lowest intermediate scale is of the
order of $10^{3}$ GeV) outside or at the periphery of the plot, away
from the mSugra point and, as the intermediate scales increase, they
converge towards the region of the mSugra point, in the middle of
the plot. In fact, note that all the blue lines of LR-class models
do touch this point, because we can slide the LR scale all the
way to $m_{G}$. But in PS- and BL- models there are two intermediate
scales and often the lowest one cannot be increased all the way up
to $m_{G}$, either because that would make the highest intermediate
scale bigger than $m_{G}$ or because it would invert the natural
ordering of the two intermediate scales. 

It is interesting to note that the BL-class with low $m_{R}$ can
produce the same imprint in the sparticle masses as LR-models. This
is to be expected because with $m_{R}$ close to $m_{B-L}$ the running
in the U(1)-mixing phase is small, leading to predictions similar
to LR-models. The equivalent limit for PS-class models is reached
for very high $m_{PS}$, close to the GUT scale (see below). On the
other hand, from fig.~(\ref{fig:LEQE_parametric_plot_RGB}) we can
see that a low $m_{PS}$ actually leads to a very different signal
on the soft sparticle masses. For example, a measurement of $LE\approx 10$
and $QE\approx 15$, together with compatible values for the other
two invariants ($QU\approx 10$ and $DL\approx -5$) would immediately 
exclude {\em all classes of models except PS-models}, and in addition 
it would strongly suggest low PS and LR scales.

\begin{figure}
\begin{centering}
\includegraphics[scale=0.25]{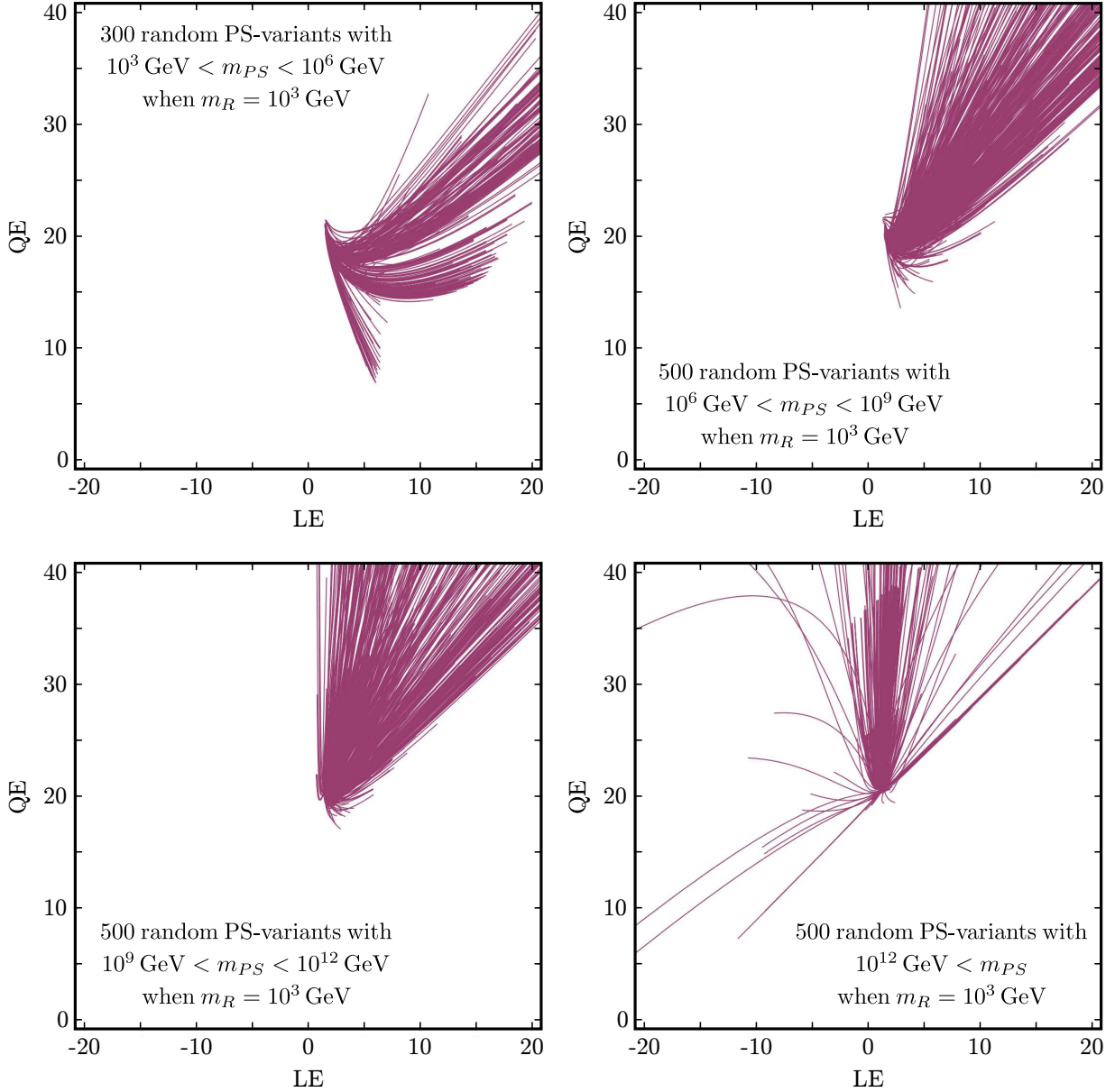}
\par\end{centering}

\caption{\label{fig:LEQE_parametric_plot_PS_only}Parametric
$\left(LE,QE\right)$ plots for different PS-variants showing the
effect of the PS scale.}
\end{figure}

Fig.~(\ref{fig:LEQE_parametric_plot_PS_only}) illustrates the general
behavior of PS-models as we increase the separation between the
$m_{LR}$ and $m_{PS}$ scales. The red region in the
$\left(LE,QE\right)$ plot tends to rotate anti-clockwise until it
reaches, for very high $m_{PS}$, the same region of points which is
predicted by LR-models.  Curiously, we also see in
fig.~(\ref{fig:LEQE_parametric_plot_PS_only}) that some of these
models actually predict different invariant values from the ones of LR
models. What happens in these cases is that since the PS phase is very
short, it is possible to have many active fields in it which decouple
at lower energies. So even though the running is short, the values of
the different gauge couplings actually get very large corrections in
this regime and these are uncommon in other settings.  For example, it
is possible in this special subclass of PS-models for $\alpha_{R}$ to
get bigger than $\alpha_{3}$/$\alpha_{4}$ before unifying!

One can see from fig. (\ref{fig:LEQE_parametric_plot_PS_only}) that 
many, although not all PS-models can lead to large values of $LE$. 
This can happen for both low and high values of $m_{PS}$ and 
is a rather particular feature of the class-II, which can 
not be found in the other classes.

\section{Summary and conclusions}

We have discussed $SO(10)$ based supersymmetric models with 
extended gauge group near the electro-weak scale, consistent 
with gauge coupling unification thanks to a ``sliding scale'' 
mechanism. We have discussed three different setups, which we 
call classes of models. The first and simplest chain we use breaks 
$SO(10)$ through a left-right symmetric stage to the SM group, 
class-II uses an additional intermediate Pati-Salam stage, while 
in class-III we discuss models which break the LR-symmetric 
group first into a $U(1)_R\times U(1)_{B-L}$ group before reaching 
the SM group. We have shown that in each case many different 
variants and many configurations (or ``proto-models'') for each 
variant can be constructed.

We have discussed that one can not only construct sliding models in
which an inverse or linear seesaw is consistent with GCU, as done in
earlier work \cite{Malinsky:2005bi,Dev:2009aw,DeRomeri:2011ie}, but
also all other known types of seesaws can, in principle, be found. We
found example configurations for seesaw type-I, type-II and type-III
and even inverse type-III (for which one example limited to class-II
was previously discussed in \cite{DeRomeri:2011ie}). 

Due to the sliding scale property the different configurations predict
potentially rich phenomenology at the LHC, although by the same
reasoning the discovery of any of the additional particles the models
predict is of course not guaranteed. However, even if all the new
particles - including the gauge bosons of the extended gauge group -
lie outside of the reach of the LHC, indirect tests of the models are
possible from measurements of SUSY particle masses. We have discussed
certain combinations of soft parameters, called ``invariants'', and
shown that the invariants themselves can be classified into a few
sets. Just determining to which set the experimental data belongs
would allow to distinguish, at least in some cases, class-I from
class-II models and also in all but one case our classes of models
are different from the ordinary high-scale seesaw (type-II and type-III)
models. Depending on the accuracy with which supersymmetric masses can
be measured in the future, the invariants could be used to gain
indirect information not only on the class of model and its variant
realized in nature, but also give hints on the scale of beyond-MSSM
physics, i.e. the energy scale at which the extended gauge group is
broken.

We add a few words of caution. First of all, our analysis is done
completely at the 1-loop level. It is known from numerical
calculations for seesaw type-II \cite{Hirsch:2008gh} and seesaw
type-III \cite{Esteves:2010ff} that the invariants receive numerically
important shifts at 2-loop level. In addition, there are also
uncertainties in the calculation from GUT-scale thresholds and from
uncertainties in the input parameters. For the latter the most
important is most likely the error on $\alpha_S$
\cite{DeRomeri:2011ie}.  With the huge number of models we have
considered, taking into account all of these effects is impractical
and, thus, our numerical results should be taken as approximate.
However, should any signs of supersymmetry be found in the future,
improvements in the calculations along these lines could be easily
made, should it become necessary.  More important for the calculation
of the invariants is, of course, the assumption that SUSY breaking
indeed is mSugra-like. Tests of the validity of this assumption can be
made also only indirectly. Many of the spectra we find, especially in
the class-II models, are actually quite different from standard mSugra
expectations and thus a pure MSSM-mSugra would give a bad fit to
experimental data, if one of these models is realized in
nature. However, all of our variants still fulfill (by construction)
a certain sum rule, see the discussion in section
\ref{sect:invclasses}.

Of course, so far no signs of supersymmetry have been seen at the LHC,
but with the planned increase of $\sqrt{s}$ for the next run of the
accelerator there is still quite a lot of parameter space to be
explored.  We note in this respect that we are not overly concerned
about the Higgs mass, $m_{h} \sim (125-126)$ GeV, if the new resonance
found by the ATLAS \cite{ATLAS:2012gk} and CMS \cite{CMS:2012gu}
collaborations turns out to be indeed the lightest Higgs boson. While
for a pure MSSM with mSugra boundary conditions it is well-known
\cite{Arbey:2011ab,Baer:2011ab,Buchmueller:2011ab,Ellis:2012aa} that
such a hefty Higgs requires multi-TeV scalars,
\footnote{Multi-TeV scalars are also required, if the MSSM with mSugra 
boundary conditions is extended to include a high-scale seesaw 
mechanism \cite{Hirsch:2012ti}.} all our models have 
an extended gauge symmetry. Thus, there are new D-terms contributing 
to the Higgs mass \cite{Haber:1986gz,Drees:1987tp}, alleviating 
the need for large soft SUSY breaking terms, as has been explicitly 
shown in \cite{Hirsch:2012kv,Hirsch:2011hg} for one particular 
realization of a class-III model \cite{Malinsky:2005bi,DeRomeri:2011ie}.

Finally, many of the configuration (or proto-models) which we have 
discussed contain exotic superfields, which might show up in the 
LHC. It might therefore be interesting to do a more detailed study 
of the phenomenology of at least some particular of the models we have 
constructed.

\section*{Acknowledgements}

This work has been supported in part by EU~Network grant UNILHC
PITN-GA-2009-237920. M.H. also 
acknowledges support from the Spanish MICINN grants
FPA2011-22975, MULTIDARK CSD2009-00064 and by the Generalitat
Valenciana grant Prometeo/2009/091. 
The work of R.M.F has been supported by
\textit{Funda\-\c{c}\~ao para a Ci\^encia e a Tecnologia} through the
fellowship SFRH/BD/47795/2008. R.M.F. and J. C. R. also acknowledge
the financial support from grants CFTP-FCT UNIT 777,
CERN/FP/123580/2011 and PTDC/FIS/102120/2008.

\appendix
\section{Lists of superfields}
\label{sec:appendix}

We have considered $SO(10)$ based models which may contain any
irreducible representation up to dimension 126 ($\boldsymbol{1}$,
$\boldsymbol{10}$, $\boldsymbol{16}$, $\overline{\boldsymbol{16}}$,
$\boldsymbol{45}$, $\boldsymbol{54}$, $\boldsymbol{120}$,
$\boldsymbol{126}$, $\overline{\mathbf{126}}$).  Once the gauge group
breaks down to $SU(4)\times SU(2)_{L}\times SU(2)_{R}$ or
$SU(3)_{C}\times SU(2)_{L}\times SU(2)_{R}\times U(1)_{B-L}$ these
$SO(10)$ fields divide into a multitude of different irreducible
representation of these groups.
In addition, if $SU(2)_{R}$ is broken
down further to $U(1)_{R}$ the following branching rules apply:
$\boldsymbol{3}\rightarrow-1,0,+1$;
$\boldsymbol{2}\rightarrow\pm\frac{1}{2}$;
$\boldsymbol{1}\rightarrow0$. The standard model's hypercharge, in the
canonical normalization, is then equal to the combination
$\sqrt{\frac{3}{5}}\left[U(1)_{R}\textrm{
hypercharge}\right]+\sqrt{\frac{2}{5}}\left[U(1)_{B-L}\textrm{
hypercharge}\right]$. In tables \ref{tab:List_of_LR_fields},
\ref{tab:List_of_PatiSalam_fields} and \ref{tab:List_of_LR_fields_U1}
we present the list of relevant fields respecting the conditions
above. In these tables we used an ordered naming of the
fields. Sometimes it is also useful, like in
table~\ref{tab:LR_field_configuration_A}, to indicate explicitly the
quantum numbers under the various groups.


\begin{table}[tbph]
\begin{center}
\scalebox{0.98}{

\begin{tabular}{c}
\begin{tabular}{lrrrrrrrrrrrrrr}
\toprule 
 & $\Phi_{1}$ & $\Phi_{2}$ & $\Phi_{3}$ & $\Phi_{4}$ & $\Phi_{5}$ & $\Phi_{6}$ & $\Phi_{7}$ & $\Phi_{8}$ & $\Phi_{9}$ & $\Phi_{10}$ & $\Phi_{11}$ & $\Phi_{12}$ & $\Phi_{13}$ & $\Phi_{14}$\tabularnewline
\cmidrule{3-15} 
 &  & $\chi$ & $\chi^{c}$ & $\Omega$ & $\Omega^{c}$ & $\Phi$ & $ $ & $ $ & $\delta_{d}$ & $\delta_{u}$ & $ $ & $ $ & $ $ & $ $\tabularnewline
\midrule
$SU(3)_{C}$ & \textbf{1} & \textbf{1} & \textbf{1} & \textbf{1} & \textbf{1} & \textbf{1} & \textbf{8} & \textbf{1} & \textbf{3} & \textbf{3} & \textbf{6} & \textbf{6} & \textbf{3} & \textbf{3}\tabularnewline
$SU(2)_{L}$ & \textbf{1} & \textbf{2} & \textbf{1} & \textbf{3} & \textbf{1} & \textbf{2} & \textbf{1} & \textbf{1} & \textbf{1} & \textbf{1} & \textbf{1} & \textbf{1} & \textbf{2} & \textbf{1}\tabularnewline
$SU(2)_{R}$ & \textbf{1} & \textbf{1} & \textbf{2} & \textbf{1} & \textbf{3} & \textbf{2} & \textbf{1} & \textbf{1} & \textbf{1} & \textbf{1} & \textbf{1} & \textbf{1} & \textbf{1} & \textbf{2}\tabularnewline
$U(1)_{B-L}$ & 0 & +1 & -1 & 0 & 0 & 0 & 0 & +2 & $-\frac{2}{3}$ & $+\frac{4}{3}$ & $+\frac{2}{3}$ & $-\frac{4}{3}$ & $+\frac{1}{3}$ & $+\frac{1}{3}$\tabularnewline
\midrule
\textbf{\footnotesize }%
\begin{tabular}{{@{}l@{}}}
PS\tabularnewline
Origin\tabularnewline
\end{tabular} & \textbf{\footnotesize }%
\begin{tabular}{{@{}r@{}}}
\textbf{\footnotesize $\Psi_{1}$}\tabularnewline
\textbf{\footnotesize $\Psi_{10}$}\tabularnewline
\end{tabular} & \textbf{\footnotesize $\overline{\Psi}_{12}$} & \textbf{\footnotesize $\Psi_{13}$} & \textbf{\footnotesize $\Psi_{3}$} & \textbf{\footnotesize $\Psi_{4}$} & \textbf{\footnotesize }%
\begin{tabular}{{@{}r@{}}}
\textbf{\footnotesize $\Psi_{2}$}\tabularnewline
\textbf{\footnotesize $\Psi_{7}$}\tabularnewline
\end{tabular} & \textbf{\footnotesize }%
\begin{tabular}{{@{}r@{}}}
\textbf{\footnotesize $\Psi_{10}$}\tabularnewline
\textbf{\footnotesize $\Psi_{11}$}\tabularnewline
\end{tabular} & \textbf{\footnotesize $\overline{\Psi}_{9}$} & \textbf{\footnotesize }%
\begin{tabular}{{@{}r@{}}}
\textbf{\footnotesize $\Psi_{8}$}\tabularnewline
\textbf{\footnotesize $\Psi_{9}$}\tabularnewline
\end{tabular} & \textbf{\footnotesize $\Psi_{10}$} & \textbf{\footnotesize $\Psi_{9}$} & \textbf{\footnotesize $\Psi_{11}$} & \textbf{\footnotesize $\Psi_{12}$} & \textbf{\footnotesize $\Psi_{13}$}\tabularnewline
\bottomrule
\end{tabular}\tabularnewline
\tabularnewline
\begin{tabular}{lrrrrrrrrrr}
\toprule 
 & $\Phi_{15}$ & $\Phi_{16}$ & $\Phi_{17}$ & $\Phi_{18}$ & $\Phi_{19}$ & $\Phi_{20}$ & $\Phi_{21}$ & $\Phi_{22}$ & $\Phi_{23}$ & $\Phi_{24}$\tabularnewline
 &  & $\Delta$ & $\Delta^{c}$ &  &  &  &  &  &  & \tabularnewline
\midrule
$SU(3)_{C}$ & \textbf{8} & \textbf{1} & \textbf{1} & \textbf{3} & \textbf{3} & \textbf{3} & \textbf{6} & \textbf{6} & \textbf{1} & \textbf{3}\tabularnewline
$SU(2)_{L}$ & \textbf{2} & \textbf{3} & \textbf{1} & \textbf{2} & \textbf{3} & \textbf{1} & \textbf{3} & \textbf{1} & \textbf{3} & \textbf{2}\tabularnewline
$SU(2)_{R}$ & \textbf{2} & \textbf{1} & \textbf{3} & \textbf{2} & \textbf{1} & \textbf{3} & \textbf{1} & \textbf{3} & \textbf{3} & \textbf{2}\tabularnewline
$U(1)_{B-L}$ & 0 & -2 & -2 & $+\frac{4}{3}$ & $-\frac{2}{3}$ & $-\frac{2}{3}$ & $+\frac{2}{3}$ & $+\frac{2}{3}$ & 0 & $-\frac{2}{3}$\tabularnewline
\midrule
\textbf{\footnotesize }%
\begin{tabular}{{@{}l@{}}}
PS\tabularnewline
Origin\tabularnewline
\end{tabular} & \textbf{\footnotesize $\Psi_{7}$} & \textbf{\footnotesize $\Psi_{16}$} & \textbf{\footnotesize $\Psi_{17}$} & \textbf{\footnotesize $\Psi_{7}$} & \textbf{\footnotesize }%
\begin{tabular}{{@{}r@{}}}
\textbf{\footnotesize $\Psi_{14}$}\tabularnewline
\textbf{\footnotesize $\Psi_{16}$}\tabularnewline
\end{tabular} & \textbf{\footnotesize }%
\begin{tabular}{{@{}r@{}}}
\textbf{\footnotesize $\Psi_{15}$}\tabularnewline
\textbf{\footnotesize $\Psi_{17}$}\tabularnewline
\end{tabular} & \textbf{\footnotesize $\Psi_{16}$} & \textbf{\footnotesize $\Psi_{17}$} & \textbf{\footnotesize $\Psi_{5}$} & \textbf{\footnotesize $\Psi_{6}$}\tabularnewline
\bottomrule
\end{tabular}\tabularnewline
\end{tabular}

}
\end{center}

\caption{\label{tab:List_of_LR_fields}Naming conventions and transformation 
properties of fields in the left-right symmetric regime (not
  considering conjugates). The charges under the $U(1)_{B-L}$ group
  shown here were multiplied by a factor $\sqrt{\frac{8}{3}}$.}
\end{table}

\begin{table}[tbph]
\begin{center}
\scalebox{1.00}{

\begin{centering}
\begin{tabular}{lrrrrrrrrrrrrrrrrr}
\toprule 
 & $\Psi_{1}$ & $\Psi_{2}$ & $\Psi_{3}$ & $\Psi_{4}$ & $\Psi_{5}$ & $\Psi_{6}$ & $\Psi_{7}$ & $\Psi_{8}$ & $\Psi_{9}$ & $\Psi_{10}$ & $\Psi_{11}$ & $\Psi_{12}$ & $\Psi_{13}$ & $\Psi_{14}$ & $\Psi_{15}$ & $\Psi_{16}$ & $\Psi_{17}$\tabularnewline
\midrule
$SU(4)$ & \textbf{1} & \textbf{1} & \textbf{1} & \textbf{1} & \textbf{1} & \textbf{6} & \textbf{15} & \textbf{6} & \textbf{10} & \textbf{15} & \textbf{20'} & \textbf{4} & \textbf{4} & \textbf{6} & \textbf{6} & \textbf{10} & \textbf{10}\tabularnewline
$SU(2)_{L}$ & \textbf{1} & \textbf{2} & \textbf{3} & \textbf{1} & \textbf{3} & \textbf{2} & \textbf{2} & \textbf{1} & \textbf{1} & \textbf{1} & \textbf{1} & \textbf{2} & \textbf{1} & \textbf{3} & \textbf{1} & \textbf{3} & \textbf{1}\tabularnewline
$SU(2)_{R}$ & \textbf{1} & \textbf{2} & \textbf{1} & \textbf{3} & \textbf{3} & \textbf{2} & \textbf{2} & \textbf{1} & \textbf{1} & \textbf{1} & \textbf{1} & \textbf{1} & \textbf{2} & \textbf{1} & \textbf{3} & \textbf{1} & \textbf{3}\tabularnewline
\midrule
\textbf{\footnotesize }%
\begin{tabular}{{@{}l@{}}}
$SO(10)$ \tabularnewline
Origin\tabularnewline
\end{tabular} & \textbf{\footnotesize }%
\begin{tabular}{{@{}r@{}}}
\textbf{\footnotesize 1}\tabularnewline
\textbf{\footnotesize 54}\tabularnewline
\end{tabular} & \textbf{\footnotesize }%
\begin{tabular}{{@{}r@{}}}
\textbf{\footnotesize 10}\tabularnewline
\textbf{\footnotesize 120}\tabularnewline
\end{tabular} & \textbf{\footnotesize 45} & \textbf{\footnotesize 45} & \textbf{\footnotesize 54} & \textbf{\footnotesize }%
\begin{tabular}{{@{}r@{}}}
\textbf{\footnotesize 45}\tabularnewline
\textbf{\footnotesize 54}\tabularnewline
\end{tabular} & \textbf{\footnotesize }%
\begin{tabular}{{@{}r@{}}}
\textbf{\footnotesize 120}\tabularnewline
\textbf{\footnotesize 126}\tabularnewline
\end{tabular} & \textbf{\footnotesize }%
\begin{tabular}{{@{}r@{}}}
\textbf{\footnotesize 10}\tabularnewline
\textbf{\footnotesize 126}\tabularnewline
\end{tabular} & \textbf{\footnotesize 120} & \textbf{\footnotesize 45} & \textbf{\footnotesize 54} & \textbf{\footnotesize 16} & \textbf{\footnotesize $\overline{\mathbf{16}}$} & \textbf{\footnotesize 120} & \textbf{\footnotesize 120} & \textbf{\footnotesize 126} & \textbf{\footnotesize $\overline{\mathbf{126}}$}\tabularnewline
\bottomrule
\end{tabular}
\par\end{centering}

}
\end{center}

\caption{\label{tab:List_of_PatiSalam_fields}
Naming conventions and transformation properties of fields in the
Pati-Salam regime (not considering conjugates)}
\end{table}

\begin{table}[tbph]
\begin{center}
\scalebox{1.0}{

\begin{tabular}{c}
\begin{tabular}{lrrrrrrrrrrrrrrrr}
\toprule 
 & $\Phi_{1}^{'}$  & $\Phi_{2}^{'}$  & $\Phi_{3}^{'}$  & $\Phi_{4}^{'}$  & $\Phi_{5}^{'}$  & $\Phi_{6}^{'}$  & $\Phi_{7}^{'}$  & $\Phi_{8}^{'}$  & $\Phi_{9}^{'}$  & $\Phi_{10}^{'}$  & $\Phi_{11}^{'}$  & $\Phi_{12}^{'}$  & $\Phi_{13}^{'}$  & $\Phi_{14}^{'}$  & $\Phi_{15}^{'}$  & $\Phi_{16}^{'}$ \tabularnewline
\midrule
$SU(3)_{C}$ & \textbf{1}  & \textbf{1}  & \textbf{1}  & 1 & \textbf{1}  & \textbf{1}  & \textbf{1}  & \textbf{8}  & \textbf{1}  & \textbf{3}  & \textbf{3}  & \textbf{6}  & \textbf{6}  & \textbf{3}  & \textbf{3} & \textbf{3}\tabularnewline
$SU(2)_{L}$ & \textbf{1}  & \textbf{2}  & \textbf{1}  & 1 & \textbf{3}  & \textbf{1}  & \textbf{2}  & \textbf{1}  & \textbf{1}  & \textbf{1}  & \textbf{1}  & \textbf{1}  & \textbf{1}  & \textbf{2}  & \textbf{1} & \textbf{1}\tabularnewline
$U(1)_{R}$ & 0  & 0  & $-\frac{1}{2}$  & $+\frac{1}{2}$ & 0  & +1  & $+\frac{1}{2}$  & 0  & 0  & 0  & 0  & 0  & 0  & 0  & $-\frac{1}{2}$ & $+\frac{1}{2}$\tabularnewline
$U(1)_{B-L}$ & 0  & +1  & -1  & -1 & 0  & 0  & 0  & 0  & +2  & $-\frac{2}{3}$  & $+\frac{4}{3}$  & $+\frac{2}{3}$  & $-\frac{4}{3}$  & $+\frac{1}{3}$  & $+\frac{1}{3}$ & $+\frac{1}{3}$\tabularnewline
\midrule
\textbf{\footnotesize }%
\begin{tabular}{{@{}l@{}}}
LR\tabularnewline
Origin\tabularnewline
\end{tabular} & {\footnotesize }%
\begin{tabular}{@{}r@{}}
\textbf{\footnotesize $\Phi_{1}$}{\footnotesize{} }\tabularnewline
\textbf{\footnotesize $\Phi_{5}$} \tabularnewline
\end{tabular} & \textbf{\footnotesize $\Phi_{2}$}{\footnotesize{} } & \textbf{\footnotesize $\Phi_{3}$}{\footnotesize{} } & \textbf{\footnotesize $\Phi_{3}$} & {\footnotesize }%
\begin{tabular}{@{}r@{}}
\textbf{\footnotesize $\Phi_{4}$}{\footnotesize{} }\tabularnewline
\textbf{\footnotesize $\Phi_{23}$} \tabularnewline
\end{tabular} & \textbf{\footnotesize $\Phi_{5}$}{\footnotesize{} } & \textbf{\footnotesize $\Phi_{6}$}{\footnotesize{} } & \textbf{\footnotesize $\Phi_{7}$}{\footnotesize{} } & {\footnotesize }%
\begin{tabular}{@{}r@{}}
\textbf{\footnotesize $\Phi_{8}$}{\footnotesize{} }\tabularnewline
\textbf{\footnotesize $\bar{\Phi}_{17}$} \tabularnewline
\end{tabular}{\footnotesize{} } & {\footnotesize }%
\begin{tabular}{@{}r@{}}
\textbf{\footnotesize $\Phi_{9}$}{\footnotesize{} }\tabularnewline
\textbf{\footnotesize $\Phi_{20}$} \tabularnewline
\end{tabular} & \textbf{\footnotesize $\Phi_{10}$}{\footnotesize{} } & {\footnotesize }%
\begin{tabular}{@{}r@{}}
\textbf{\footnotesize $\Phi_{11}$}{\footnotesize{} }\tabularnewline
\textbf{\footnotesize $\Phi_{22}$} \tabularnewline
\end{tabular}{\footnotesize{} } & \textbf{\footnotesize $\Phi_{12}$}{\footnotesize{} } & \textbf{\footnotesize $\Phi_{13}$}{\footnotesize{} } & \textbf{\footnotesize $\Phi_{14}$} & \textbf{\footnotesize $\Phi_{14}$}\tabularnewline
\bottomrule
\end{tabular}\tabularnewline
\tabularnewline
\begin{tabular}{lrrrrrrrrrrrrrrr}
\toprule 
 & $\Phi_{17}^{'}$  & $\Phi_{18}^{'}$  & $\Phi_{19}^{'}$  & $\Phi_{20}^{'}$  & $\Phi_{21}^{'}$  & $\Phi_{22}^{'}$  & $\Phi_{23}^{'}$  & $\Phi_{24}^{'}$  & $\Phi_{25}^{'}$  & $\Phi_{26}^{'}$  & $\Phi_{27}^{'}$  & $\Phi_{28}^{'}$  & $\Phi_{29}^{'}$  & $\Phi_{30}^{'}$  & $\Phi_{31}^{'}$\tabularnewline
\midrule
$SU(3)_{C}$ & \textbf{8}  & \textbf{1}  & \textbf{1}  & \textbf{1}  & \textbf{3}  & \textbf{3}  & \textbf{3}  & \textbf{3}  & \textbf{3}  & \textbf{6}  & \textbf{6}  & \textbf{6}  & \textbf{1}  & \textbf{3} & \textbf{3}\tabularnewline
$SU(2)_{L}$ & \textbf{2}  & \textbf{3}  & \textbf{1}  & \textbf{1}  & \textbf{2}  & \textbf{2}  & \textbf{3}  & \textbf{1}  & \textbf{1}  & \textbf{3}  & \textbf{1}  & \textbf{1}  & \textbf{3}  & \textbf{2} & \textbf{2}\tabularnewline
$U(1)_{R}$ & $+\frac{1}{2}$  & 0  & -1  & +1  & $-\frac{1}{2}$  & $+\frac{1}{2}$  & 0  & -1  & +1  & 0  & -1  & +1  & +1  & $-\frac{1}{2}$ & $+\frac{1}{2}$\tabularnewline
$U(1)_{B-L}$ & 0  & -2  & -2  & -2  & $+\frac{4}{3}$  & $+\frac{4}{3}$  & $-\frac{2}{3}$  & $-\frac{2}{3}$  & $-\frac{2}{3}$  & $+\frac{2}{3}$  & $+\frac{2}{3}$  & $+\frac{2}{3}$  & 0  & $-\frac{2}{3}$ & $-\frac{2}{3}$\tabularnewline
\midrule
\textbf{\footnotesize }%
\begin{tabular}{{@{}l@{}}}
LR\tabularnewline
Origin\tabularnewline
\end{tabular} & \textbf{\footnotesize $\Phi_{15}$}{\footnotesize{} } & \textbf{\footnotesize $\Phi_{16}$}{\footnotesize{} } & \textbf{\footnotesize $\Phi_{17}$}{\footnotesize{} } & \textbf{\footnotesize $\Phi_{17}$}{\footnotesize{} } & \textbf{\footnotesize $\Phi_{18}$}{\footnotesize{} } & \textbf{\footnotesize $\Phi_{18}$}{\footnotesize{} } & \textbf{\footnotesize $\Phi_{19}$}{\footnotesize{} } & \textbf{\footnotesize $\Phi_{20}$}{\footnotesize{} } & \textbf{\footnotesize $\Phi_{20}$}{\footnotesize{} } & \textbf{\footnotesize $\Phi_{21}$}{\footnotesize{} } & \textbf{\footnotesize $\Phi_{22}$}{\footnotesize{} } & \textbf{\footnotesize $\Phi_{22}$}{\footnotesize{} } & \textbf{\footnotesize $\Phi_{23}$}{\footnotesize{} } & \textbf{\footnotesize $\Phi_{24}$} & \textbf{\footnotesize $\Phi_{24}$}\tabularnewline
\bottomrule
\end{tabular}\tabularnewline
\end{tabular}

}
\end{center}

\caption{\label{tab:List_of_LR_fields_U1}
Naming conventions and transformation properties of fields in the U(1)
mixing regime (not considering conjugates).  The charges under the
$U(1)_{B-L}$ group shown here were multiplied by a factor
$\sqrt{\frac{8}{3}}$.}
\end{table}

In order for a group $G$ to break into a subgroup $H\subset G$,
there must be a field transforming non-trivially under $G$ which
contains a singlet of $H$ that acquires vacuum expectation value.
From this observation alone we know that certain fields must be present
in a fundamental model if we are to achieve a given breaking sequence:
\begin{itemize}
\item The breaking $PS\rightarrow LR$ is possible only with the (\textbf{15},\textbf{1},\textbf{1})
while $PS\rightarrow3211$ requires the combination (\textbf{15},\textbf{1},\textbf{1})
+ (\textbf{1},\textbf{1},\textbf{3}). For the direct breaking $PS\rightarrow321$
there are two choices: (\textbf{4},\textbf{1},\textbf{2}), (\textbf{10},\textbf{1},\textbf{3})
or their conjugates;
\item The breaking $LR\rightarrow3211$ requires the (\textbf{1},\textbf{1},\textbf{3},0)
representation while the direct route $LR\rightarrow321$ is possible
with the presence of (\textbf{1},\textbf{1},\textbf{2},-1), (\textbf{1},\textbf{1},\textbf{3},-2)
or their conjugates;
\item The group $3211$ can be broken down to $321$ with the representations
$(\boldsymbol{1},\boldsymbol{1},\frac{1}{2},-1)$, $(\boldsymbol{1},\boldsymbol{1},1,-2)$
or their conjugates.
\end{itemize}


\begin{thebibliography}{10}

\bibitem{Fukuda:1998mi}
Super-Kamiokande Collaboration, Y.~Fukuda {\em et~al.},
\newblock Phys.Rev.Lett. {\bf 81}, 1562 (1998), arXiv:hep-ex/9807003.

\bibitem{Ahmad:2002jz}
SNO Collaboration, Q.~Ahmad {\em et~al.},
\newblock Phys.Rev.Lett. {\bf 89}, 011301 (2002), arXiv:nucl-ex/0204008.

\bibitem{Eguchi:2002dm}
KamLAND Collaboration, K.~Eguchi {\em et~al.},
\newblock Phys.Rev.Lett. {\bf 90}, 021802 (2003), arXiv:hep-ex/0212021.

\bibitem{Tortola:2012te}
D.~Forero, M.~Tortola, and J.W.F.~Valle,
\newblock Phys.Rev. {\bf D86}, 073012 (2012), arXiv:1205.4018.

\bibitem{Minkowski:1977sc}
P.~Minkowski,
\newblock Phys.Lett. {\bf B67}, 421 (1977).

\bibitem{Yanagida:1979ss}
T.~Yanagida,
\newblock KEK lectures, ed. O.~Sawada and A.~Sugamoto, Tsukuba, Japan  (1979).

\bibitem{GellMann:1980vs}
M.~Gell-Mann, P.~Ramond, and R.~Slansky,
\newblock Conf.Proc. {\bf C790927}, 315 (1979).

\bibitem{Mohapatra:1979ia}
R.~N. Mohapatra and G.~Senjanovic,
\newblock Phys.Rev.Lett. {\bf 44}, 912 (1980).

\bibitem{Fritzsch:1974nn}
H.~Fritzsch and P.~Minkowski,
\newblock Annals Phys. {\bf 93}, 193 (1975).

\bibitem{Mohapatra:1986uf}
R.~N.~Mohapatra,
\newblock Contemporary Physics (Springer, Berlin, Germany)  (1986).

\bibitem{Cvetic:1983su}
M.~Cvetic and J.~C. Pati,
\newblock Phys.Lett. {\bf B135}, 57 (1984).

\bibitem{Kuchimanchi:1993jg}
R.~Kuchimanchi and R.~Mohapatra,
\newblock Phys.Rev. {\bf D48}, 4352 (1993), arXiv:hep-ph/9306290.

\bibitem{Aulakh:1997ba}
C.~S. Aulakh, K.~Benakli, and G.~Senjanovic,
\newblock Phys.Rev.Lett. {\bf 79}, 2188 (1997), arXiv:hep-ph/9703434.

\bibitem{Aulakh:1997fq}
C.~S. Aulakh, A.~Melfo, A.~Rasin, and G.~Senjanovic,
\newblock Phys.Rev. {\bf D58}, 115007 (1998), arXiv:hep-ph/9712551.

\bibitem{Majee:2007uv}
S.~K. Majee, M.~K. Parida, A.~Raychaudhuri, and U.~Sarkar,
\newblock Phys.Rev. {\bf D75}, 075003 (2007), arXiv:hep-ph/0701109.

\bibitem{DeRomeri:2011ie}
V.~De~Romeri, M.~Hirsch, and M.~Malinsky,
\newblock Phys.Rev. {\bf D84}, 053012 (2011), arXiv:1107.3412.

\bibitem{Calibbi:2009cp}
L.~Calibbi, L.~Ferretti, A.~Romanino, and R.~Ziegler,
\newblock Phys.Lett. {\bf B672}, 152 (2009), arXiv:0812.0342.

\bibitem{Malinsky:2005bi}
M.~Malinsky, J.~Romao, and J.W.F.~Valle,
\newblock Phys.Rev.Lett. {\bf 95}, 161801 (2005), arXiv:hep-ph/0506296.

\bibitem{Dev:2009aw}
P.~B. Dev and R.~Mohapatra,
\newblock Phys.Rev. {\bf D81}, 013001 (2010), arXiv:0910.3924.

\bibitem{Pati:1974yy}
J.~C. Pati and A.~Salam,
\newblock Phys.Rev. {\bf D10}, 275 (1974).

\bibitem{Buckley:2006nv}
M.~R. Buckley and H.~Murayama,
\newblock Phys.Rev.Lett. {\bf 97}, 231801 (2006), arXiv:hep-ph/0606088.

\bibitem{Hirsch:2008gh}
M.~Hirsch, S.~Kaneko, and W.~Porod,
\newblock Phys.Rev. {\bf D78}, 093004 (2008), arXiv:0806.3361.

\bibitem{Esteves:2010ff}
J.~Esteves, J.~Romao, M.~Hirsch, F.~Staub, and W.~Porod,
\newblock Phys.Rev. {\bf D83}, 013003 (2011), arXiv:1010.6000.

\bibitem{Beringer:1900zz}
Particle Data Group, J.~Beringer {\em et~al.},
\newblock Phys.Rev. {\bf D86}, 010001 (2012).

\bibitem{wp}
J.~Romao {\em et~al.},
\newblock http://porthos.ist.utl.pt/arXiv/AllSO10GUTs.

\bibitem{mohapatra:1986bd}
R.~N. Mohapatra and J.~W.~F. Valle,
\newblock Phys. Rev. {\bf D34}, 1642 (1986).

\bibitem{Akhmedov:1995ip}
E.~K. Akhmedov, M.~Lindner, E.~Schnapka, and J.W.F.~Valle,
\newblock Phys.Lett. {\bf B368}, 270 (1996), arXiv:hep-ph/9507275.

\bibitem{Akhmedov:1995vm}
E.~K. Akhmedov, M.~Lindner, E.~Schnapka, and J.W.F.~Valle,
\newblock Phys.Rev. {\bf D53}, 2752 (1996), arXiv:hep-ph/9509255.

\bibitem{Foot:1988aq}
R.~Foot, H.~Lew, X.~He, and G.~C. Joshi,
\newblock Z.Phys. {\bf C44}, 441 (1989).

\bibitem{Babu:1998tm}
K.~Babu, B.~Dutta, and R.~Mohapatra,
\newblock Phys.Rev. {\bf D60}, 095004 (1999), arXiv:hep-ph/9812421.

\bibitem{Davidson:1993qk}
S.~Davidson, D.~C. Bailey, and B.~A. Campbell,
\newblock Z.Phys. {\bf C61}, 613 (1994), arXiv:hep-ph/9309310.

\bibitem{Kuznetsov:2012ad}
A.~Kuznetsov, N.~Mikheev, and A.~Serghienko,
\newblock (2012), arXiv:1210.3697.

\bibitem{Fonseca:2011vn}
R.~M. Fonseca, M.~Malinsky, W.~Porod, and F.~Staub,
\newblock Nucl.Phys. {\bf B854}, 28 (2012), arXiv:1107.2670.

\bibitem{Baer:2012vr}
H.~Baer, V.~Barger, A.~Lessa, and X.~Tata,
\newblock (2012), arXiv:1207.4846.

\bibitem{ATLAS-CONF-2012-109}
ATLAS Collaboration, G.~Aad {\em et~al.},
\newblock ATLAS-CONF-2012-109  (2012).

\bibitem{ATLAS:2012gk}
ATLAS Collaboration, G.~Aad {\em et~al.},
\newblock Phys.Lett. {\bf B716}, 1 (2012), arXiv:1207.7214.

\bibitem{CMS:2012gu}
CMS Collaboration, S.~Chatrchyan {\em et~al.},
\newblock Phys.Lett. {\bf B716}, 30 (2012), arXiv:1207.7235.

\bibitem{Arbey:2011ab}
A.~Arbey, M.~Battaglia, A.~Djouadi, F.~Mahmoudi, and J.~Quevillon,
\newblock Phys.Lett. {\bf B708}, 162 (2012), arXiv:1112.3028.

\bibitem{Baer:2011ab}
H.~Baer, V.~Barger, and A.~Mustafayev,
\newblock Phys.Rev. {\bf D85}, 075010 (2012), arXiv:1112.3017.

\bibitem{Buchmueller:2011ab}
O.~Buchmueller {\em et~al.},
\newblock Eur.Phys.J. {\bf C72}, 2020 (2012), arXiv:1112.3564.

\bibitem{Ellis:2012aa}
J.~Ellis and K.~A. Olive,
\newblock Eur.Phys.J. {\bf C72}, 2005 (2012), arXiv:1202.3262.

\bibitem{Hirsch:2012ti}
M.~Hirsch, F.~Joaquim, and A.~Vicente,
\newblock JHEP {\bf 1211}, 105 (2012), arXiv:1207.6635.

\bibitem{Haber:1986gz}
H.~E. Haber and M.~Sher,
\newblock Phys.Rev. {\bf D35}, 2206 (1987).

\bibitem{Drees:1987tp}
M.~Drees,
\newblock Phys.Rev. {\bf D35}, 2910 (1987).

\bibitem{Hirsch:2012kv}
M.~Hirsch, W.~Porod, L.~Reichert, and F.~Staub,
\newblock Phys.Rev. {\bf D86}, 093018 (2012), arXiv:1206.3516.

\bibitem{Hirsch:2011hg}
M.~Hirsch, M.~Malinsky, W.~Porod, L.~Reichert, and F.~Staub,
\newblock JHEP {\bf 1202}, 084 (2012), arXiv:1110.3037.

\end{thebibliography}
\end{document}